\documentclass[]{jfm}

\usepackage{graphicx}
\usepackage{newtxtext}
\usepackage{newtxmath}
\usepackage{natbib}
\usepackage{hyperref}
\hypersetup{
    colorlinks = true,
    urlcolor   = blue,
    citecolor  = black,
}

\newcommand{\RomanNumeralCaps}[1]
\linenumbers

\usepackage{color,amsmath,amssymb,verbatim,tabularx}

\title{Radiation-driven melting and porous flow in the near-surface layer of an ice sheet}

\author{Tilly Woods\aff{1,2}
  \corresp{\email{tilly.woods@lmu.de}},
 \and Ian J. Hewitt\aff{1}}

\affiliation{\aff{1}Mathematical Institute, University of Oxford, Oxford OX2 6GG, UK
\aff{2}Ludwig-Maximilians-Universität, Munich, Germany}

\usepackage{natbib}

\renewenvironment{thebibliography}[1]{
\begin{oldthebibliography}{#1}
\setlength{\itemsep}{0.0cm}
\setlength{\parskip}{0.0cm}}
{\end{oldthebibliography} }

\newcommand{\be}{\begin{equation} }
\newcommand{\ee}{\end{equation}}
\newcommand{\bmath}{\begin{displaymath} }
\newcommand{\emath}{\end{displaymath} }

\newcommand{\dd}[2]{\frac{\textrm{d}#1}{\textrm{d} #2}}
\newcommand{\pd}[2]{\frac{\partial #1}{\partial #2}}

\newcommand{\sub}[1]{_{\mathrm{#1}}}

\begin{document}
\maketitle

\begin{abstract}
The ablation area of an ice sheet can undergo melting both directly at the surface and in the metre or so below the surface, due to the penetration and absorption of shortwave radiation.  The fate of the meltwater has important implications for the mass balance of the ice sheet; some of it runs off into surface streams and eventually to the ocean, while some is stored in the porous `weathering crust', and can refreeze.  We present a fluid-mechanical model to analyse the downward percolation and lateral drainage of the meltwater in the crust, and its influence on porosity, saturation, temperature and melting rate.  Previous models have either ignored the possibility for lateral drainage, in which case the crust is always saturated with water, or have made ad-hoc assumptions about where and how the meltwater runs off.  Through a series of analytical and numerical solutions to the model, we find that lateral drainage has the effect of controlling the water table within the weathering crust, which separates an unsaturated region near the surface from a saturated aquifer beneath.  Under fluctuating weather conditions that induce melting and refreezing, the lateral drainage can result in the near-surface ice being more porous and also colder than is the case without lateral drainage.  The model is able to predict the partitioning of the meltwater between surface and subsurface runoff, and refreezing, and has potential to be incorporated within larger-scale models of the ice-sheet mass balance.
\end{abstract}

% \begin{keywords}
% Authors should not enter keywords on the manuscript, as these must be chosen by the author during the online submission process and will then be added during the typesetting process (see \href{https://www.cambridge.org/core/journals/journal-of-fluid-mechanics/information/list-of-keywords}{Keyword PDF} for the full list).  Other classifications will be added at the same time.
% \end{keywords}

\section{Introduction}

    The loss of ice from ice sheets and glaciers is controlled in large part by the rate at which meltwater is produced and then runs off from their surface.  In regions where the surface is bare of snow, shortwave solar radiation is partially absorbed at the ice surface and partially penetrates below the surface, causing both surface and subsurface heating and melting.  This creates a porous near-surface layer referred to as a `weathering crust' \citep{Muller1969,Cooper2018}.  Meltwater can be transported through and transiently stored in this porous layer, affecting how much water runs off into the ocean \citep[e.g.][]{Mikkelsen2016,Noel2017}.  Understanding the interaction of thermodynamics and fluid mechanics that controls these processes is therefore important.  

    Both radiation-driven melting and porous meltwater flow have been considered in previous models of ice-sheet surface melt, but separately.
    The subsurface penetration of shortwave radiation has been included in models of melt ponds on sea ice \citep[e.g.][]{Taylor2004}, lakes on the surface of ice sheets \citep[e.g.][]{Buzzard2018AShelf,Law2020}, and the porous weathering crust \citep{Schuster2001,Hoffman2014,Woods2023,Woods2024}. 
    The radiation is most simply modelled as decaying exponentially with depth according to Beer's law \citep[e.g.][]{Bear2018}, with the rate of decay being described by an extinction coefficient. Another common technique is to use a two-stream approximation \citep[e.g.][]{Schlatter1972,Perovich1990,Liston1999,Taylor2004,Taylor2005}, decomposing the net radiation into downward and upward travelling components which satisfy a pair of coupled differential equations. The two-stream approximation has the benefit of enabling optical properties to depend on evolving ice properties, such as porosity and water content. In both approaches, the radiation can either be modelled spectrally (considering a continuum of wavelengths with wavelength-dependent optical properties) \citep[e.g.][]{Liston1999,Hoffman2014} or using a broadband approach (considering a single wavelength-independent value of shortwave radiation) \citep[e.g.][]{Taylor2004,Taylor2005,Woods2023}.

    Porous meltwater flow has been considered in models of (vertical) percolation through firn (partially-compacted snow) \citep[e.g.][]{Colbeck1972,Gray1996,Ligtenberg2011,Meyer2017,Moure2023} and the lateral transport and storage of meltwater in firn aquifers \citep{Miller2020,Miller2022}. The majority of percolation models are one-dimensional, and many are inherently numerical, with water being transferred from one discrete layer to the next using a `tipping bucket' scheme \citep[e.g.][]{Ligtenberg2011,KuipersMunneke2014,KuipersMunneke2015}. Other more mathematical treatments make use of Darcy's law to describe flow driven by gravity and capillary pressure \citep[e.g.][]{Colbeck1972,Colbeck1974,Colbeck1976,Gray1996}.
    % Darcy's law is used in the model presented in the thesis and will be introduced later in this chapter.
    Some of these percolation models are isothermal \citep[e.g.][]{Colbeck1972,Colbeck1974,Sellers2000}, but others have also considered phase change \citep[melting and refreezing, e.g.][]{Colbeck1976,Meyer2017,Moure2023}.

\begin{figure}
    \centering
    \includegraphics[width=\textwidth]{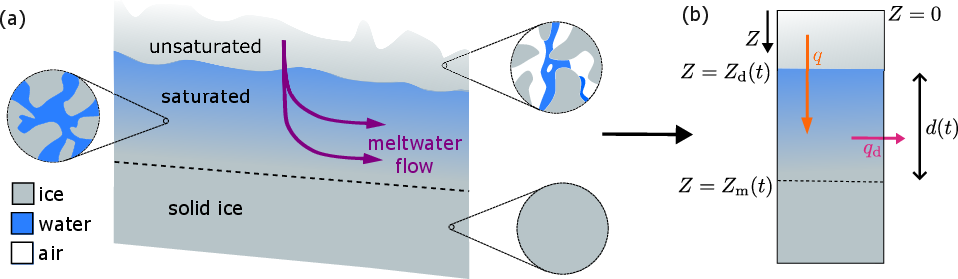}
    \caption{(a) A schematic of the flow of meltwater through the weathering crust, and (b) how we represent this in the one-dimensional model, using the downward water flux $q$ and a parametrisation $q_{\mathrm{d}}$ for the lateral drainage. The thickness of the saturated region is $d = Z_{\mathrm{m}} - Z_{\mathrm{d}}$, where $Z=Z_{\mathrm{d}}$ is the position of the water table and $Z=Z_{\mathrm{m}}$ is the bottom of the weathering crust, with $Z$ measured vertically downwards from the ice surface.}
    \label{fig4:WC_water_table_schematic_3D_1D}
\end{figure}

    Our goal in this study is to capture both radiation-driven melt and porous meltwater flow in a single continuum model for the weathering crust.
    % We make use of a broadband Beer's law for the penetrating radiation and Darcy's law for meltwater flow. 
    % The model is an extension of the model presented in \citet{Woods2023,Woods2024}, with the crucial difference that we allow for lateral drainage of the meltwater, and consequently also vertical percolation and partial saturation of the pore space.
    The weathering crust is a transient layer of porous ice, on the order of a metre in depth, with impermeable ice below \citep{Muller1969,Cooper2018}. 
    It is formed by radiation-driven subsurface melting \citep{Brandt1993,Cooper2021}, and evolves in response to changing weather conditions \citep{Schuster2001,Tedstone2020}. 
    Most of the weathering crust is saturated with meltwater \citep{Cooper2018,Cook2016,Irvine-Fynn2021}, above which there is a few centimetres of highly-porous, unsaturated ice (containing both air and water), as shown schematically in figure~\ref{fig4:WC_water_table_schematic_3D_1D}(a). 
    We ignore the possibility of any overlying snow, which may be present in the early melt season.

    Meltwater flows laterally through the weathering crust with typical flow speeds of $<0.13$ m/day \citep{Irvine-Fynn2021}. This plays an important role in transporting meltwater, microbes and nutrients across the ice sheet \citep[][]{Stibal2012,Cook2016,Cooper2018}, and the carbon fluxes between the ice sheet and the atmosphere \citep{Cook2012}.
    Drainage of meltwater through the crust also impacts the albedo (reflectivity) of the surface \citep{Tedstone2020}, which is a crucial control on how much the surface melts. For example, if a weathering crust forms on a wet, bare ice surface, the surface water is able to drain away through the weathering crust, significantly changing the appearance of the ice surface (increasing the albedo).

    Existing models of the weathering crust \citep{Schuster2001,Hoffman2014,Woods2023,Woods2024} are all one-dimensional (vertical) and include radiation penetration but do not model the flow of meltwater. 
    \citet{Hoffman2014} did consider the effect of lateral meltwater drainage in their one-dimensional model by removing any meltwater exceeding a 10\% melt fraction at each timestep, but they did not model any (vertical) meltwater flow or density/porosity evolution explicitly. Their work did however demonstrate the importance of accounting for lateral drainage, even in a one-dimensional model; without lateral drainage, observed surface lowering, porosity and temperature were not correctly captured \citep{Hoffman2014}. Moreover, \citet{Stevens2026} found that the \citet{Woods2024} model (without meltwater flow) overestimated the ice density compared to ice-core observations of weathering crusts on the Greenland Ice Sheet. They suggest that the overestimation is due to too much meltwater being refrozen, and that this problem could be addressed by introducing a meltwater flow model.

    In this study, we extend the one-dimensional model presented in \citet{Woods2023} to include meltwater flow.  We investigate the impact of lateral drainage, which crucially allows for vertical percolation of meltwater and for the crust to be only partially saturated.  We make use of a broadband Beer's law for the penetrating radiation and Darcy's law for meltwater flow, driven by gravity and capillary forces. We account for lateral drainage in the one-dimensional model using a parametrisation based on Darcy's law.
    % \tw{Meltwater flow not naturally possible (except transiently) in 1D model because of impermeable base? Or in main paper?}

    In Section~\ref{sec:model}, we present and then non-dimensionalise the model. In Sections~\ref{sec4:isothermal_water} and~\ref{sec4:steadily_melting}, we apply the model to two test cases for which we can compare numerical and semi-analytical solutions: isothermal water transport and steadily melting solutions. In Section~\ref{sec:periodic_switching}, we model an idealised `periodic switching' scenario that captures several of the key weather-related behaviours observed in reality. Finally, in Section~\ref{sec:conclusion}, we summarise and discuss the insights gained.

\section{Model}
\label{sec:model}

% \begin{itemize}
%     \item 1D unsat model with lateral drainage.

%     \item Maybe repeat the full 3D model from the 2023 paper too?
% \end{itemize}

% \begin{itemize}
%     \item Extend (3D) WC model presented in \citet{Woods2023}.

%     \item Add Darcy's law and lateral drainage

%     \item Won't repeat full 3D model here but will give 1D reduction (which is what we'll use)

%     \item present 1D equations

%     \item enthalpy formulation

%     \item non-dimensionalisation - present both dim and non-dim in main paper, or one in just appendix?

% \end{itemize}

Our model accounts for mass conservation of water and ice, energy conservation, and surface and sub-surface absorption of radiation.  It is an extension of the model used in \citet{Woods2023,Woods2024} which, due to being purely one-dimensional and assuming impermeable ice below, involved a fully-saturated weathering crust and no meltwater flow.  Here, we continue to work with a one-dimensional model, but now parameterise lateral drainage, which allows for the crust to become unsaturated and for vertical percolation.  The setup is shown in figure~\ref{fig4:WC_water_table_schematic_3D_1D}(b). 

We consider porous ice with the pores containing water and/or air. The volume fraction of ice is $1-\phi$, where $\phi$ is the porosity. The saturation $S$ gives the proportion of the pore space that contains water. Hence the total volume fraction of water is $\phi S$ and the total volume fraction of air is $\phi (1-S)$.
We assume local thermal equilibrium, meaning that the ice-water-air mixture is at the melting temperature whenever ice and water are both present. We also assume for simplicity that the ice and water have shared material and thermal properties, and that the air density is negligible.

We write the equations in a coordinate system in which $Z = h(t)-z$ measures distance downwards from the surface $z = h(t)$. Since the surface is in general lowering, this induces advective terms $(\dot{h})$ in the evolution equations. The conversion to this moving coordinate system from the original Eulerian coordinate system is described in \citet{Woods2023,Woods2024}.

\subsection{Darcy's law}
Flow through a porous medium driven by gravity and capillary forces, is described by Darcy's law \citep[e.g.][]{Bear2018}. The vertical volume flux of water per unit area is given by Darcy's law as
\be
\label{eq4:Darcy1D}
q = \frac{k(\phi) k_{\mathrm{rw}} (S)}{\mu} \left( \rho g -\pd{p_{\mathrm{w}}}{Z} \right),
\ee
where $p_{\mathrm{w}}$ is the water pressure, $k(\phi)$ is the (absolute) permeability, $k_{\mathrm{rw}} (S)$ is the relative permeability, $\mu$ is the viscosity of water, and $g$ is the acceleration due to gravity. We parametrise the permeability and relative permeability in terms of the porosity $\phi$ and saturation $S$, respectively, as
\be
k(\phi)= k_0 \phi^3, \quad k_{\mathrm{rw}}(S) = S^{3}.
\ee
We choose these forms to make it possible to find analytical solutions, but note that other parametrisations are also commonly used \citep[e.g.][]{Kozeny1927,Carman1937,Fowler2011}. 

In unsaturated regions ($0<S<1$), the water pressure is parametrised in terms of the saturation $S$ as
\be
\label{eq4:pw}
p_{\mathrm{w}} = -p_c(S) = - \frac{\gamma}{d_p} S^{- 2},
\ee
where $\gamma$ is the surface tension and $d_p$ is the mean grain size. Here, we have followed the parameterisation of \citet{Meyer2017} for the capillary pressure $p\sub{c}$, which relates the air and water pressures, $p\sub{a}$ and $p\sub{w}$, as $p\sub{a} - p\sub{w} = p\sub{c}(S)$. We have also used the assumption that air pressure is negligible and constant to set $p\sub{a} = 0$ without loss of generality. 

In saturated regions ($S=1$), $p_{\mathrm{w}}$ is not related directly to $S$ but instead satisfies an elliptic equation that arises from the constraint of incompressibility (see below).

\subsection{Lateral drainage}
% The impermeable base of the weathering crust means that vertical flow in a one-dimensional model is no longer possible once the crust becomes saturated; there is nowhere for the water to flow to. In reality, there is significant drainage of water via the lateral flow through the weathering crust. 
To incorporate the lateral flow of water into our one-dimensional model, we use a parametrisation for lateral drainage based on Darcy's law, giving the lateral drainage flux $q_{\mathrm{d}}$ as
\be
\label{eq4:qd_dim}
q_d =  \frac{\rho g}{\mu} k_{\mathrm{h}}(\phi) k_{\mathrm{rw}} (S) \sin\theta.
\ee
This represents downslope, gravity-driven flow through a weathering crust with surface slope $\sin\theta$, where $\theta$ is the (small) angle between the horizontal and the ice surface, and
\be
k_{\mathrm{h}}(\phi) = \beta_{\mathrm{h}} k (\phi)
\ee
is the horizontal permeability.
% , related to the vertical permeability $k(\phi)$ by an anisotropy factor $\beta_{\mathrm{h}}$. 
The permeability of the weathering crust is anisotropic, with structural features such as horizontal, impermeable ice layers making it easier for water to flow horizontally than vertically \citep{Cooper2018}. We account for this in our model using the anisotropy factor $\beta_{\mathrm{h}}\ge 1$. 
%Since the surface slope is typically small \citep[e.g. $\sin\theta = 0.022$][]{Irvine-Fynn2021}, the horizontal and downslope directions are approximately equal, and it is appropriate to use the \emph{downslope} component of gravity with the \emph{horizontal} permeability in the lateral flow parametrisation \eqref{eq4:qd_dim}.

The lateral drainage $q_{\mathrm{d}}$ is assumed to occur over a lengthscale $L_{\mathrm{d}}$, which could represent the distance between supraglacial streams into which the water drains. This results in a `sink' term $-q_{\mathrm{d}}/L_{\mathrm{d}}$ in the one-dimensional model below.

%We also note that in a purely one-dimensional model (without a lateral drainage parametrisation), the impermeable base of the weathering crust means that vertical flow is only possible transiently. Once the pore space becomes fully saturated, there is nowhere for the water to flow (other than running off at the surface).

\subsection{Mass conservation}

Mass conservation for ice and water is expressed in the one-dimensional model as
\be
\label{eq4:mass_ice_unsat_1d}
\rho \bigg( \pd{}{t}(1-\phi) + \dot{h} \pd{}{Z} (1-\phi) \bigg) = -m_{\mathrm{int}},
\ee
\be
\label{eq4:mass_water_unsat_1d}
\rho \bigg( \pd{}{t} (\phi S) + \dot{h} \pd{}{Z} (\phi S) + \pd{q}{Z} \bigg) = m_{\mathrm{int}} - \rho \frac{q_{\mathrm{d}}}{L_{\mathrm{d}}},
\ee
where $\rho$ is the density of ice and water (assumed here to be equal), $\dot{h}$ is the rate of change of surface height, $m\sub{int}$ is the subsurface melt rate, $ q_{\mathrm{d}}$ is the lateral drainage~\eqref{eq4:qd_dim} over lengthscale $L_{\mathrm{d}}$, and $q$ is the downward water flux given by Darcy's law~\eqref{eq4:Darcy1D}.

\subsection{Energy conservation}
Conservation of energy is expressed as
\be
\label{eq4:energy_eq_unsat}
\begin{split}
\rho c \bigg( \pd{}{t}\big( (1-\phi +\phi S)T \big) + \dot{h}  \pd{}{Z} \big((1-\phi &+\phi S)T \big) + \pd{}{Z}(qT) \bigg) \\ 
= k \pd{}{Z}\bigg( (1-\phi +\phi S) \pd{T}{Z} \bigg) 
&- \pd{F}{Z} - \mathcal{L} m_{\mathrm{int}} - \rho c \frac{q_{\mathrm{d}} T}{L_{\mathrm{d}}},
\end{split}
\ee
where $T$ is the temperature of the ice-water mixture, $k$ is the thermal conductivity, $c$ is the specific heat capacity, and $F$ is net downward shortwave radiation. The final term represents the heat removed from the system by the lateral drainage. When the mixture is at the melting temperature ($T=T\sub{m}$), equation \eqref{eq4:energy_eq_unsat} is an equation for the subsurface melt rate $m\sub{int}$. Otherwise, there is no subsurface melting ($m\sub{int}=0$) and equation \eqref{eq4:energy_eq_unsat} is an equation for the temperature $T$.

%\subsection{Radiation model}

% The net downward subsurface shortwave radiation $F$ in \citet{Woods2023} is modelled using the two-stream approximation, which allows for absorption and scattering to depend on porosity and saturation, for example. Here, we simplify by assuming that absorption and scattering are constant and uniform throughout the weathering crust. Hence the solution to the two-stream approximation is \citep{Woods2023,Woods2024}
The net downward shortwave radiation $F$ is modelled as 
\be
\label{eq:F}
F = (1-\chi) (1-a) Q\sub{si} e^{- \kappa Z},
\ee
where $\chi$ is the proportion of absorbed shortwave radiation that is absorbed \emph{at} the surface (see below), $a$ is the albedo, $Q\sub{si}$ is the incoming shortwave radiation, and $\kappa$ is the extinction coefficient. 
As an alternative to (\ref{eq:F}), it would be relatively straightforward to use a more general two-stream approximation, which would allow for absorption and scattering to be made dependent on porosity and saturation and to account for consequent feedbacks; see \citet{Woods2023}. 

\subsection{Boundary conditions}

The boundary conditions for the mass and energy conservation equations are as follows.
The surface energy balance
\begin{equation}
\label{eq4:SEB_unsat}
\begin{split}
k (1- \phi + & \phi S) \frac{\partial T}{\partial Z} + \chi (1-a) Q_{\mathrm{si}} + Q_0\\
&- \big(\upsilon + \rho cR_{\mathrm{r}} \big) (T - T_{\mathrm{m}}) 
= \rho \mathcal{L} m_{\mathrm{surf}} \quad \mathrm{at} \quad Z=0,
\end{split}
\end{equation}
demands that the energy fluxes at the surface must either balance, or, if not, any excess energy is used for melting the surface. Either the surface is at the melting temperature $T=T\sub{m}$ and the surface energy balance \eqref{eq4:SEB_unsat} determines the surface melt rate $m\sub{surf}$, or the surface is below the melting temperature, $T<T\sub{m}$, there is no surface melting $m\sub{surf}=0$, and the surface energy balance \eqref{eq4:SEB_unsat} is a boundary condition for the energy equation (\ref{eq4:energy_eq_unsat}). 

In equation \eqref{eq4:SEB_unsat}, the first term is the conductive heat flux from the weathering crust, $\chi (1-a) Q\sub{si}$ is the shortwave radiation absorbed at the surface, $Q_0-\upsilon(T-T\sub{m})$ is a linear representation of the net incoming longwave radiation and turbulent heat fluxes (with expressions for $Q_0$ and $\upsilon$ given in \citet{Woods2023}), and $\rho c R\sub{r}(T\sub{m}-T)$ is the energy input from rainfall at rate $R\sub{r}$.  We have assumed here that rain falls at the melting temperature. We neglect the accumulation of snowfall (and its associated energy input), but note that snowfall could be incorporated by making some slight modifications \citep[e.g.][]{Meyer2017}.

The surface mass balance is expressed by
\be
\label{eq4:kinematic_ice_unsat}
\rho (1-\phi) \dot{h}= -\rho m_{\mathrm{surf}} \quad \mathrm{at} \quad Z=0,
\ee
and
\be
\label{eq4:kinematic_water_unsat}
\rho \big(q + \phi S \dot{h} \big) = \rho (m_{\mathrm{surf}} + R_{\mathrm{r}} - r_{\mathrm{off}}) \quad \mathrm{at} \quad Z=0,
\ee
where $r\sub{off}$ is the surface runoff rate (the rate at which water runs off over the surface; this will be combined with the subsurface lateral drainage to give the \emph{total} runoff rate).

When the surface is unsaturated ($S<1$ at $Z=0$), we assume that any surface water will percolate down into the pore space. Hence the surface runoff~$r_{\mathrm{off}}$ is zero, and~\eqref{eq4:kinematic_water_unsat} provides a boundary condition relating $q$ and $S$ at $Z=0$. When the surface is saturated ($S=1$ at $Z=0$), we assume that excess surface water is removed instantaneously as surface runoff, with $r_{\mathrm{off}}$ given by~\eqref{eq4:kinematic_water_unsat}. In this case, we prescribe the boundary condition $p_{\mathrm{w}}=0$ at $Z=0$, which allows us to solve for the water pressure $p_{\mathrm{w}}$ and hence water flux $q$ in the saturated region (see appendix for more details). 

Finally, we prescribe the temperature $T_{\infty} < T\sub{m}$ deep in the ice, which we assume to have zero porosity,
\be
\label{eq4:Tinf_BC_dim}
T \to T_{\infty} 
,\quad \phi \to 0 
\quad \mathrm{as} \quad Z \to \infty.
\ee
% \be
% \label{eq4:phiinf_BC_dim}
% \phi \to 0 \quad \mathrm{as} \quad Z \to \infty.
% \ee

\subsection{Nondimensionalisation}

For the rest of this manuscript, we choose to work with dimensionless variables. This motivates neglecting certain small terms in our equations, making analytical solutions possible (see Section~\ref{sec4:isothermal_water}).

We non-dimensionalise by writing $T=T_{\mathrm{m}} + [T] \hat{T}$ (where $[\cdot]$ is used to denote the `scale') and scaling all other variables with
\be
\label{eq4:weathering_scalings}
\begin{split}
[T] = & \frac{[t][F]}{\rho c [Z]}, \quad [m_{\mathrm{int}}] = \frac{\rho}{[t]}, \quad [m_{\mathrm{surf}}]=[r_{\mathrm{off}}]= \frac{[Z]}{[t]}, \\
%[H] = \rho c [T], \quad 
& [q] = \frac{[Z]}{[t]}, \quad [p_{\mathrm{w}}] = \rho g [Z], \quad [h]=[Z],
\end{split}
\ee
with $[Z]$, $[t]$ and $[F]$ prescribed. Typically, we use $[Z] = 1$ m, $[F] = 200$ W m$^{-2}$, and $[t]=1$ day or 1 year. Typical values of the physical parameters and scalings are shown in tables~\ref{tab:parameters_dimensional} and~\ref{tab4:scalings}, respectively.

The following dimensionless parameters are thus introduced:
\be
\label{eq4:weathering_crust_dimensionless_parameters}
\begin{split}
\hat{k} = \frac{k[t]}{\rho c [Z]^2}, \quad 
& \mathcal{S} = \frac{\mathcal{L}}{c [T]}, \quad \hat{\kappa} = \kappa [Z], \\
\mathcal{K} = \frac{k_0 \rho g [t]}{\mu [Z]},  \quad \mathcal{B} = & \frac{\rho g [Z] d_p}{\gamma},  \quad \mathcal{K}_{\mathrm{d}} = \frac{\sin\theta\, \beta_{\mathrm{h}} [Z]}{L_{\mathrm{d}}},\\
\end{split}
\ee
where $\hat{k}$ is the dimensionless thermal conductivity, $\mathcal{S}$ is the Stefan number (the ratio of latent heat to sensible heat, with a larger latent heat meaning more energy is required to achieve the same amount of melting), 
%$\hat{\alpha}$ and $\hat{r}$ are dimensionless absorption and scattering coefficients, 
$\hat{\kappa}$ is the dimensionless extinction coefficient, $\mathcal{K}$ is the dimensionless hydraulic conductivity (a measure of how easily water can flow through the porous ice), $\mathcal{B}$ is the Bond number (a measure of the effect of gravity compared to capillary pressure), and $\mathcal{K}_{\mathrm{d}}$ is the dimensionless lateral drainage strength/efficiency. $\mathcal{K}_{\mathrm{d}}$ accounts for the effect of surface slope and the anisotropy of the permeability. The dimensionless forcings are
\be
\mathcal{Q}_{\mathrm{si}} = \frac{Q_{\mathrm{si}}}{[F]}, \quad \mathcal{Q}_0 = \frac{Q_0}{[F]}, \quad \hat{\upsilon} = \frac{\upsilon [T]}{[F]}, \quad \mathcal{T}_{\infty} = \frac{T_{\infty} - T_{\mathrm{m}}}{[T]}, \quad \mathcal{R}\sub{r} = \frac{R\sub{r} [t]}{[Z]}.
\ee
Typical values of the dimensionless parameters and forcings are shown in table~\ref{tab4:dimensionless_parameters}. We now summarise the dimensionless model, from now on dropping the hats from the dimensionless variables.

\begin{table}
\centering
\begin{tabularx}{\textwidth}{|X|c|c|c|}
\hline
Quantity & Symbol & Default value & Units\\
\hline
Density of ice & $\rho$ & 910 & kg m$^{-3}$ \\

Specific heat capacity of ice at 0$^{\mathrm{o}}$C &$c$ & 2097 & J kg$^{-1}$ K$^{-1}$ \\

Thermal conductivity& $k$ & 2.1 &  kg m$^{-3}$ K$^{-1}$\\

Melting temperature& $T_{\mathrm{m}}$ & 273.15 & K  \\

Latent heat of melting &$\mathcal{L}$ & 334000 & m$^2$ s$^{-2}$  \\

% Heat transfer coefficient$^{\mathrm{a}}$ & $\tilde{C}$ & $2.58 \times 10^{-5}$ & \\

% Air pressure$^{\mathrm{a}}$ & $p_{\mathrm{a}}$ & $8 \times 10^{4}$ & Pa \\

% Wind speed$^{\mathrm{a}}$ & $u$ & 5 & m s$^{-1}$ \\

% Surface emissivity$^{\mathrm{b}}$ & $\epsilon$ & 0.97 & \\

% Stefan-Boltzmann constant & $\sigma$ & $5.67 \times 10^{-8}$ & W m$^{-2}$ K$^{-1}$ \\

Albedo & $a$ & 0.6 & \\

Proportion of absorbed shortwave radiation that is absorbed at the surface & $\chi$  & 0.36&  \\

Extinction coefficient of subsurface shortwave radiation$^{\mathrm{b}}$ & $\kappa$ & 1.5  & m$^{-1}$ \\

% Unnamed optical property$^{\mathrm{e}}$ & $s$ &  0.7009 & \\

% Absorption coefficient$^{\mathrm{e}}$ &$\alpha$ &  0.2637  & m$^{-1}$ \\

% Scattering coefficient$^{\mathrm{e}}$ & $r$ & 4.1338  & m$^{-1}$ \\

Mean grain size$^{\mathrm{a}}$ & $d_{\mathrm{p}}$ & $10^{-4}$ & m \\

Surface tension$^{\mathrm{a}}$ & $\gamma$ & 0.07 & N m$^{-1}$ \\

Permeability constant$^{\mathrm{b}}$ & $k_0$ & $5.56 \times 10^{-11}$ & m$^2$\\

Viscosity of water & $\mu$ & $10^{-3}$ & Pa s \\

\hline

Incoming shortwave radiation flux & $Q_{\mathrm{si}}$ & various  & W m$^{-2}$ \\

Turbulent heat and longwave radiation flux & $Q_0$ & various  & W m$^{-2}$ \\

Effective heat transfer coefficient & $\upsilon$ & 14.8 & W m$^{-2}$ K$^{-1}$ \\

Temperature at depth & $T_{\infty}$ & -10  & $^{\mathrm{o}}$C \\

Rainfall rate & $R\sub{r}$ & 0 & m s$^{-1}$\\

\hline

\end{tabularx}
\caption{Default parameter values. These are the same as the default values in \citet{Woods2024}, but with the addition of $d\sub{p}$, $\gamma$, $k_0$ and $\mu$. 
% $^{\mathrm{a}}$Motivated by \citet{Hoffman2014}. $^{\mathrm{b}}$\citet{Taylor2005}.
Quantities below the line are model forcings.
$^{\mathrm{a}}$\citet{Meyer2017}.
$^{\mathrm{b}}$Calculated as $k_0 = d_{\mathrm{p}}^2 /180$ \citep[][]{Fowler2011}. 
% $^{\mathrm{e}}$\citet{Woods2024}. Many of the ice properties are taken from \citet{Cuffey2010}.
}
\label{tab:parameters_dimensional}
\end{table}

\begin{table}
    \centering
    \begin{tabular}{|c|c|}
    \hline
    Scaling & Typical value \\
    \hline
        $[t]$ & 1 day \\
        $[Z],[h]$  & 1 m \\
        $[F]$  & 200 W m$^{-2}$ \\
        $[T]$ & 9.05 K \\
        $[m_{\mathrm{surf}}],[r_{\mathrm{off}}],[q]$ & 1 m d$^{-1}$ \\
      %  $[H]$ & 1.7 $\times$10$^{7}$ kg m$^{-1}$ s$^{-2}$ \\
        $[p_{\mathrm{w}}]$ & 8.9 kPa \\
        $[m_{\mathrm{int}}]$ & 910 kg m$^{-3}$ d$^{-1}$ \\
        % $[q]$ & $[Z]/[t]$ & 1 m d$^{-1}$\\
        % $[h]$ & 1 m\\
        % $[W],[\phi],[S]$ & 1 \\
        \hline
    \end{tabular}
    \caption{Scalings used to non-dimensionalise the model, with typical values calculated from the values shown in table~\ref{tab:parameters_dimensional}. Note that $\phi$ and $S$ are already dimensionless so are not listed here.}
    \label{tab4:scalings}
\end{table}

\begin{table}
    \centering
    \begin{tabularx}{\textwidth}{|c|c|X|}
    \hline
    Symbol & Typical value & Dimensionless parameter \\
    \hline
        $\mathcal{S}$ & 17.6 & Stefan number\\%[1ex]
        $\hat{k}$  & 0.095 &  Dimensionless thermal conductivity \\%[1ex]
        % $a$  & 0.6 & Albedo \\%[1ex]
        % $\chi$  & 0.36& Proportion of absorbed shortwave radiation absorbed at the surface \\%[1ex]
        $\hat{\kappa}$  & 1.5 & Dimensionless extinction coefficient \\%[1ex]
        $\mathcal{K}$  & 42.9 & Hydraulic conductivity scale\\%[1ex]
        $\mathcal{B}$  & $12.8$ & Bond number \\%[1ex]
        $\mathcal{K}_{\mathrm{d}}$  & 0.01\;--\;100 & Lateral drainage strength/efficiency \\
        \hline
        $\mathcal{Q}_{\mathrm{si}}$ & 0\;--\;5 & Dimensionless shortwave radiation \\%[1ex]

        $\mathcal{Q}_0$  & -3\;--\;2 & Dimensionless longwave radiation and turbulent heat fluxes \\%[1ex]

        $\hat{\upsilon}$  & 0.67 & Dimensionless effective heat transfer coefficient \\%[1ex]

        $\mathcal{T}_{\infty}$  & -1.1 & Dimensionless far-field temperature\\%[1ex]

        % $\mathcal{R}\sub{s}$  & 0 & Dimensionless snowfall rate\\%[1ex]

        $\mathcal{R}\sub{r}$  & 0\;--\;0.1 & Dimensionless rainfall rate\\%[1ex]
        \hline
    \end{tabularx}
    \caption{Dimensionless parameters in the weathering crust model. Typical values are shown, calculated using the parameters values and scalings shown in table~\ref{tab:parameters_dimensional} and table~\ref{tab4:scalings}. 
    Parameters below the line are the dimensionless model forcings.}
    %$^{*}$The Bond number is assumed to be infinite in Sections~\ref{sec4:isothermal_water} and \ref{sec4:steadily_melting}.
    \label{tab4:dimensionless_parameters}
\end{table}

The dimensionless mass and energy conservation equations are
\be
\label{eq4:mass_ice_unsat_nondim}
\pd{}{t}(1-\phi) + \dot{h} \pd{}{Z} (1-\phi)  = -m_{\mathrm{int}},
\ee
\be
\label{eq4:mass_water_unsat_nondim}
\pd{}{t} (\phi S) + \dot{h} \pd{}{Z} (\phi S) + \pd{q}{Z}  = m_{\mathrm{int}} - q_{\mathrm{d}},
\ee
\be
\label{eq4:temp_eqn_unsat_nondim}
\pd{}{t}\big( W T \big) + \dot{h}  \pd{}{Z} \big(WT \big) + \pd{}{Z}(qT) 
= \hat{k} \pd{}{Z}\bigg( W \pd{T}{Z} \bigg) - \pd{F}{Z} - \mathcal{S} m_{\mathrm{int}} - q_{\mathrm{d}} T,
\ee
where we have written $W = 1-\phi+\phi S$, and the net downward subsurface radiation \eqref{eq:F} is
\be
\label{eq:F_nondim}
F = (1-\chi)(1-a) \mathcal{Q}\sub{si} e^{-\hat{\kappa} Z}.
\ee

% The dimensionless equations used in the enthalpy method (equivalent to equations~\eqref{eq4:mass_water_unsat_nondim} and~\eqref{eq4:temp_eqn_unsat_nondim}) are
% \be
% \label{eq4:Weqn_nondim}
% \pd{W}{t} + \pd{}{Z} \left( \dot{h} W + q \right) =  - q_{\mathrm{d}},
% \ee
% \be
% \label{eq4:Heqn_nondim}
% \pd{H}{t} + \pd{}{Z} \left( \dot{h} H -\hat{k} W \pd{T}{Z} + F + (\mathcal{S} + T) q \right) =  - (\mathcal{S}+T) q_{\mathrm{d}},
% \ee

The lateral drainage flux~\eqref{eq4:qd_dim} is
\be
% q_d = \mathcal{K}_{\mathrm{d}} \mathcal{K}  k(\phi) k_{\mathrm{rw}} (S),
q_d = \mathcal{K}_{\mathrm{d}} \mathcal{K}  \phi^3 S^3,
\ee
and the downward water flux~\eqref{eq4:Darcy1D} is
\be
\label{eq4:Darcy_nondim}
% q = \mathcal{K} k(\phi) k_{\mathrm{rw}} (S) \left( 1 - \pd{p_{\mathrm{w}}}{Z} \right),
q = \mathcal{K} \phi^3 S^3 \left( 1 - \pd{p_{\mathrm{w}}}{Z} \right).
\ee
% with
% \be
% k(\phi) = \phi^3, \quad k_{rw}(S) = S^3.
% \ee
Either $S=1$ and $p_{\mathrm{w}}$ must be solved for, or $0<S<1$ and $p_{\mathrm{w}}$ is parametrised in terms of $S$ as 
\be
\label{eq4:pw_nondim}
p_{\mathrm{w}} = - \frac{1}{\mathcal{B}} S^{-2}.
\ee

The dimensionless boundary conditions are
\be
\label{eq4:SEB_nondim}
\hat{k} W\pd{T}{Z} + \chi (1-a) \mathcal{Q}_{\mathrm{si}} + \mathcal{Q}_0 - (\hat{\upsilon} + R_{\mathrm{r}}) T = \mathcal{S} m_{\mathrm{surf}} \quad \mathrm{at} \quad Z=0,
\ee
\be
\label{eq4:kinematic_water_nondim}
q + \phi S \dot{h} =  m_{\mathrm{surf}} + R_{\mathrm{r}} - r_{\mathrm{off}}
\quad \mathrm{at} \quad Z=0,
\ee
\be
\label{eq4:kinematic_ice_nondim}
 (1-\phi) \dot{h} = - m_{\mathrm{surf}} \quad \mathrm{at} \quad Z=0,
\ee
\be
\label{eq4:far_field_T_nondim}
T \to \mathcal{T}_{\infty} 
,\quad \phi \to 0
\quad \mathrm{as} \quad Z \to \infty.
\ee

\subsection{Model summary}
In summary, the mass and energy conservation equations \eqref{eq4:mass_ice_unsat_nondim}-\eqref{eq4:temp_eqn_unsat_nondim} 
are solved with boundary conditions \eqref{eq4:SEB_nondim}-\eqref{eq4:far_field_T_nondim} to determine the porosity $\phi$, saturation $S$, and temperature $T$, the surface melt rate $m\sub{surf}$, subsurface melt rate $m\sub{int}$, the surface runoff $r\sub{off}$ and the rate of change of surface height $\dot{h}$. This solution also involves solving for the water flux $q$ and the water pressure $p_w$ using \eqref{eq4:Darcy_nondim}-\eqref{eq4:pw_nondim},
noting that the way this happens is different in saturated and unsaturated regions.  
The numerical method used is described in appendix \ref{sec4:numerical_method}.

The prescribed `forcings' are the incoming shortwave radiation $\mathcal{Q}\sub{si}$, the longwave radiation and turbulent heat flux contributions from $\mathcal{Q}_0$ and $\hat{\upsilon}$, the rainfall $\mathcal{R}\sub{r}$ and the far-field temperature $\mathcal{T}_{\infty}$. In general, all the forcings could be time dependent, and varying weather conditions would be reflected in time-varying $\hat{\upsilon}$, $\mathcal{R}\sub{r}$, $\mathcal{Q}\sub{si}$ and $\mathcal{Q}_0$. However, in this idealised study, we take $\hat{\upsilon}$, $\mathcal{R}\sub{r}$, $\mathcal{T}_{\infty}$ to be constant and consider temporal variations of only $\mathcal{Q}\sub{si}$ and $\mathcal{Q}_0$.

\section{Isothermal water transport}
\label{sec4:isothermal_water}

To start, we consider some idealised test cases for which we can find analytical solutions. We compare these to our numerical solutions as a test of the numerical method.  They are also useful to help understand the parametric controls on the dynamics.

One test case is to consider the infiltration of (rain)water into temperate (i.e. at the melting temperature) porous ice. This removes the thermal aspect of the problem, allowing us to isolate and test the water flow. We assume that the porous ice is at the melting temperature everywhere for all time, so that both water and ice can be present. We also assume that there is no melting (subsurface or surface) or freezing, meaning that the porosity $\phi$ is constant, so $\phi = \phi(Z)$. These conditions are achieved by setting $T=0$ initially and having no prescribed radiation. We force the model with prescribed constant rainfall $\mathcal{R}_{\mathrm{r}}$ at the surface, and assume that there is no water present initially ($S=0$).

The porosity in the weathering crust is observed to decrease roughly exponentially with depth, and we therefore prescribe the porosity to be 
\be
\label{eq4:phi}
\phi (Z) = \phi_0 \exp \left(-{Z}/{Z_0} \right),
\ee
for some constants $\phi_0$ and $Z_0$.

% \begin{figure}
%     \centering
%     \includegraphics[width=\textwidth]{Figures/iso_snapshots_Kd0.eps}
%     \caption{\textcolor{red}{Maybe combine Figures \ref{fig4:iso_snapshots_Kd0} and \ref{fig4:iso_snapshots_Kd2_10} (and maybe \ref{fig4:isothermal_analytical_numerical_evolution}).} Isothermal water transport without lateral drainage ($\mathcal{K}_{\mathrm{d}} = 0$). Analytical (dashed lines) and numerical (solid lines) solutions for saturation $S$ (blue) and water flux $q$ (orange), with the fixed porosity profile shown in grey. Note that $q$ is multiplied by 10 to make it more visible in the plots. Other parameters are $\mathcal{K}=100$, $\phi_0 = 0.2$, $Z_0=0.5$, $q_0 = 0.1$, with $\mathcal{B}=10^{4}$ in the numerical simulations. Solutions shown are dimensionless.
%     % \textcolor{blue}{\texttt{isothermal\_snapshots\_3\_Kd\_vals\_fig.m}}
%     }
%     \label{fig4:iso_snapshots_Kd0}
% \end{figure}

\begin{figure}
    \centering
    \includegraphics[width=\textwidth]{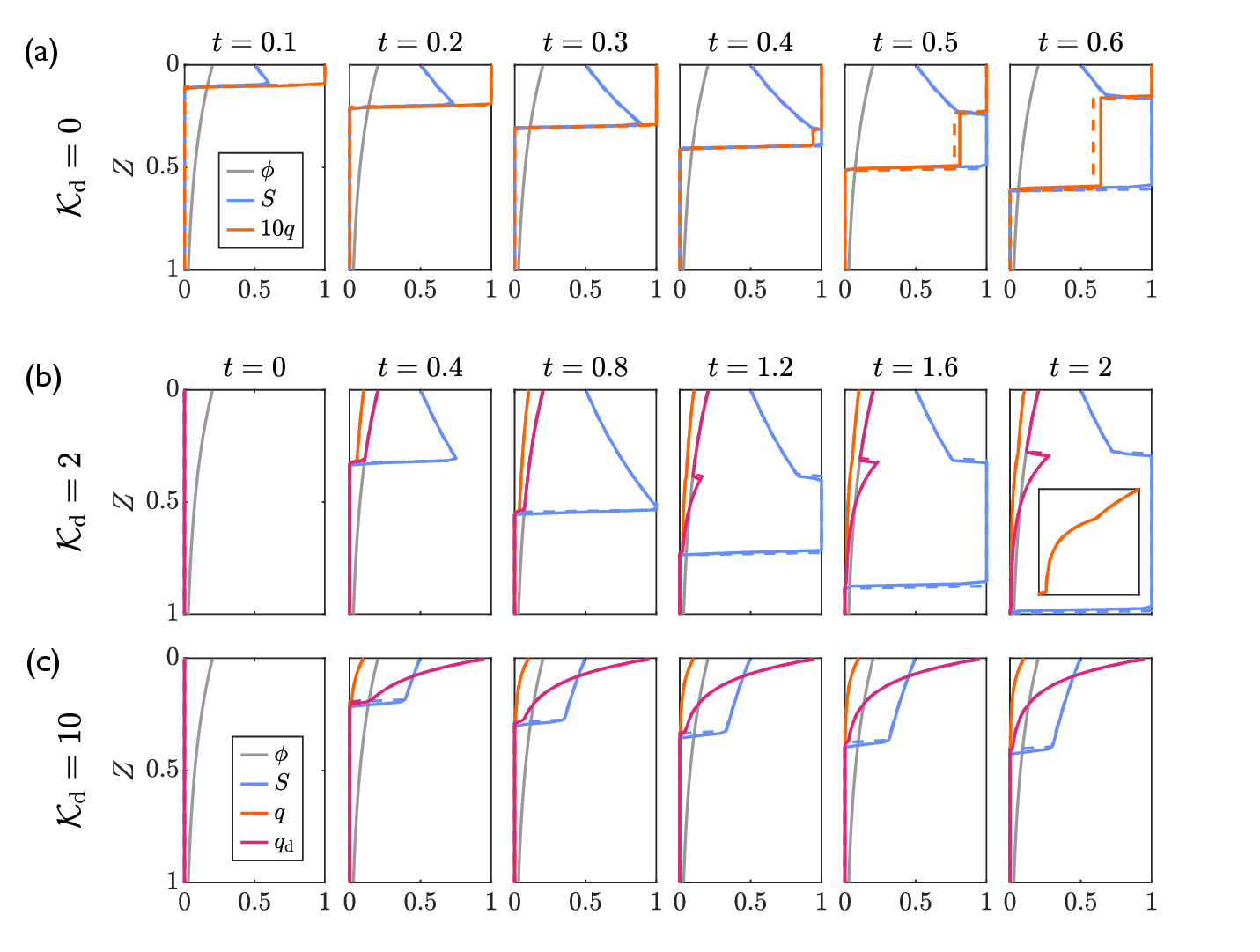}
    \caption{Isothermal water transport (a) without lateral drainage ($\mathcal{K}_{\mathrm{d}} = 0$), and (b,c) with lateral drainage ($\mathcal{K}_{\mathrm{d}} = 2$ and $\mathcal{K}_{\mathrm{d}} = 10$, respectively). Analytical (dashed lines) and numerical (solid lines) solutions for saturation $S$ (blue) and water flux $q$ (orange), and lateral drainage $q_{\mathrm{d}}$ (pink), with the fixed porosity profile shown in grey. Note that $q$ is multiplied by 10 in (a) to make it more visible. 
    %Other parameters are $\mathcal{K}=100$, $\phi_0 = 0.2$, $Z_0=0.5$, $q_0 = 0.1$, with $\mathcal{B}=10^{4}$ in the numerical simulations. Solutions shown are dimensionless.
%    The inset demonstrates the discontinuities in the gradient of $q$ across the saturation fronts.
    Other parameters are $\mathcal{K}=100$, $\phi_0 = 0.2$, $Z_0=0.5$, $\mathcal{R}_{\mathrm{r}} = 0.1$, with $\mathcal{B}=10^{4}$ in the numerical simulations. Solutions shown are dimensionless.
    }
    \label{fig4:iso_snapshots_Kd2_10}
\end{figure}

\subsection{Approximate solution neglecting capillary pressure}
\label{sec:isothermal_solution}
In order to find a semi-analytical solution,
we assume that the Bond number is large, meaning that gravity dominates capillary pressure. The method is adapted from that in \citet{Meyer2017}, which addresses the case of no lateral drainage. Since the $1/{\mathcal{B}}$ term in the flux gives rise to a diffusion term in the saturation equation, neglecting it means that the semi-analytical solution has discontinuities in $S$, which are smoothed out in the numerical solution.

With $\mathcal{B} \gg 1$, the water pressure in the unsaturated regions is approximately zero (i.e. equal to atmospheric pressure). This means that the downward water flux $q$~\eqref{eq4:Darcy_nondim} is gravity-driven, and the lateral drainage $q_{\mathrm{d}}$ is therefore proportional to $q$, as
\be
q_{\mathrm{d}} = \mathcal{K}_{\mathrm{d}} q
\ee
in unsaturated regions.
The water content equation~\eqref{eq4:mass_water_unsat_nondim} in unsaturated regions, $0<S<1$, can therefore be expressed as
\be
\label{eq4:unsaturated}
\pd{}{t}(\phi S) + \pd{q}{Z} = -\mathcal{K}_{\mathrm{d}} q \quad \mathrm{with} \quad q = \mathcal{K} \phi^3 S^3.
\ee
In saturated regions, where $S=1$, we have $q_{\mathrm{d}} = \mathcal{K}_{\mathrm{d}} \mathcal{K} \phi^3$, and therefore
\be
\label{eq4:saturated}
\pd{q}{Z} = -\mathcal{K}_{\mathrm{d}} \mathcal{K} \phi^3 \quad \mathrm{with} \quad q = \mathcal{K} \phi^3 \left( 1- \pd{p_{\mathrm{w}}}{Z} \right),
\ee
with $p_{\mathrm{w}}$ to be determined. In~\eqref{eq4:saturated}, we have used that $\phi S$ is time independent (since $S=1$ and $\phi = \phi(Z)$). The initial and boundary conditions are
\be
\label{eq4:isothermal_initial}
S = 0 \quad \mathrm{at} \quad t=0,
\ee
\be
\label{eq4:isothermal_boundary}
q = \mathcal{R}_{\mathrm{r}} \quad \mathrm{at} \quad Z=0.
\ee

\begin{figure}
    \centering
    \includegraphics[width=\textwidth]{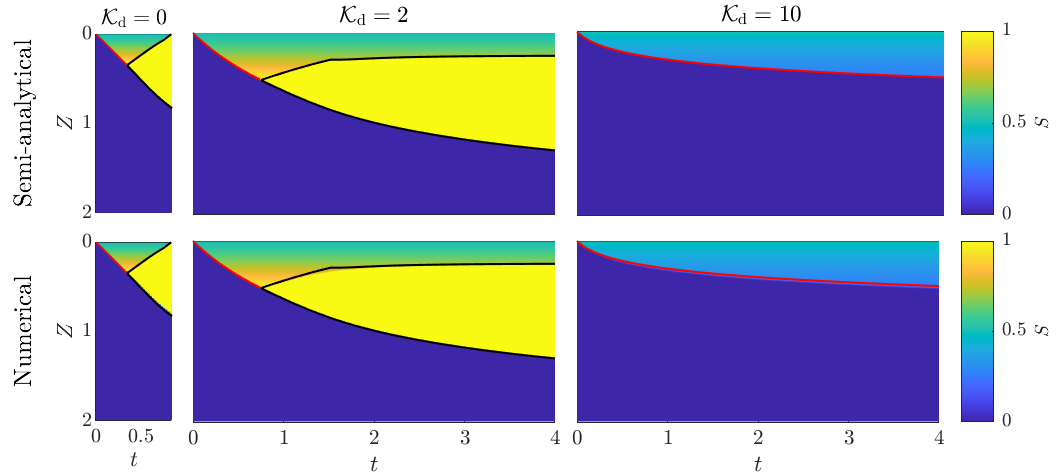}
    \caption{Semi-analytical (top row) and numerical (bottom row) solutions for the saturation $S$ in the isothermal water transport problem for 3 different values of lateral drainage strength $\mathcal{K}_{\mathrm{d}}$. The semi-analytical wetting front $Z_{\mathrm{w}}(t)$ (red) and semi-analytical saturation fronts $Z_{\mathrm{t}}(t)$, $Z_{\mathrm{b}}(t)$ (black) are overlaid. Parameter values: $\mathcal{K}=100$, $\phi_0 = 0.2$, $Z_0=0.5$, $\mathcal{R}_{\mathrm{r}} = 0.1$, with $\mathcal{B}=10^{4}$ in the numerical simulations. Solutions shown are dimensionless.
    }
\label{fig4:isothermal_analytical_numerical_evolution}
\end{figure}

We solve this problem using the method of characteristics, by first rewriting equation \eqref{eq4:unsaturated} in terms of the new variable $\tilde{W} = \phi S$. The details are given in appendix~\ref{app:isothermal_water}. Here, we just present the solution. The semi-analytical and numerical solutions for different values of lateral drainage are shown in figures~ \ref{fig4:iso_snapshots_Kd2_10} and~\ref{fig4:isothermal_analytical_numerical_evolution}, and agree well.

The influx of water at the surface into the initially dry porous ice results in a wetting front $Z_{\mathrm{w}} (t)$ propagating downwards, behind which there is unsaturated flow (figure~\ref{fig4:iso_snapshots_Kd2_10}(a) until $t\approx 0.4$; (b) until $t\approx 0.8$ and (c) for all time). The solutions for the saturation $S$ and water flux $q$ are
\be
\label{eq4:S_ana_time_evol}
S =
\begin{cases}
    \frac{1}{\phi_0} \big( \frac{\mathcal{R}_{\mathrm{r}}}{\mathcal{K}} \big)^{1/3} e^{(1/Z_0-\mathcal{K}_{\mathrm{d}}/3)Z}, & Z<Z_{\mathrm{w}}(t), \\
    0, & Z>Z_{\mathrm{w}}(t),
\end{cases}
\ee
and
\be
\label{eq4:q_ana_time_evol}
q =
\begin{cases}
    \mathcal{R}_{\mathrm{r}} e^{-\mathcal{K}_{\mathrm{d}} Z}, & Z<Z_{\mathrm{w}}(t), \\
    0, & Z>Z_{\mathrm{w}}(t),
\end{cases}
\ee
where the wetting front is given by
\be
\label{eq4:Zw_drainage}
Z_{\mathrm{w}}(t) = \frac{3}{2 \mathcal{K}_{\mathrm{d}}} \log \Big(1 + \frac{2}{3} \mathcal{K}_{\mathrm{d}} \mathcal{K}^{1/3} \mathcal{R}_{\mathrm{r}}^{2/3} t \Big).
\ee
This solution holds until saturation ($S=1$) occurs.  For sufficiently large values of lateral drainage ($\mathcal{K}_{\mathrm{d}}>3/Z_0$) this never occurs, since the saturation decreases with depth (figure~\ref{fig4:iso_snapshots_Kd2_10}, $\mathcal{K}_{\mathrm{d}}=10$).  But for sufficiently small lateral drainage ($\mathcal{K}_{\mathrm{d}}<3/Z_0$), the saturation increases with depth, and after some time $t\sub{s}$, saturation occurs at the wetting front (figure~\ref{fig4:iso_snapshots_Kd2_10}(a), $t\approx 0.4$; (b) $t\approx 0.8$). The depth $Z\sub{s}$ at which saturation first occurs is given by
\be
\label{eq4:Zs_drainage}
Z_{\mathrm{s}} = \frac{Z_0}{3 - Z_0 \mathcal{K}_{\mathrm{d}}} \log \Bigg( \frac{\mathcal{K} \phi_0^3 }{\mathcal{R}_{\mathrm{r}}} \Bigg),
\ee
and the time $t_{\mathrm{s}}$ is given by
\be
\label{eq4:ts_drainage}
t_{\mathrm{s}} = \frac{3}{2 \mathcal{K}_{\mathrm{d}} \mathcal{K}^{1/3} \mathcal{R}_{\mathrm{r}}^{2/3}} \Bigg( \bigg( \frac{\mathcal{K} \phi_0^3}{\mathcal{R}_{\mathrm{r}}} \bigg)^{\frac{2 Z_0 \mathcal{K}_{\mathrm{d}}}{9 - 3 Z_0 \mathcal{K}_{\mathrm{d}}}} - 1 \Bigg).
\ee
% For this solution to hold, we must have $\mathcal{K}_{\mathrm{d}}<3/Z_0$. That is, the lateral drainage must not be too large.
For $t>t\sub{s}$, there is a growing saturated region $Z_{\mathrm{t}}(t)<Z<Z_{\mathrm{b}}(t)$ that grows both upwards and downwards.
%(figure~\ref{fig4:iso_snapshots_Kd0}, $t=0.4$ onwards; figure~\ref{fig4:iso_snapshots_Kd2_10}, $t=0.8$ onwards). 
The solution for $t>t\sub{s}$ is
\be
\label{eq4:S_ana_time_evol_sat}
S =
\begin{cases}
    \frac{1}{\phi_0} \big( \frac{\mathcal{R}_{\mathrm{r}}}{\mathcal{K}} \big)^{1/3} e^{(1/Z_0-\mathcal{K}_{\mathrm{d}}/3)Z}, & Z<Z_{\mathrm{t}}(t), \\
    1, & Z_{\mathrm{t}}(t)<Z<Z_{\mathrm{b}}(t), \\
    0, & Z>Z_{\mathrm{b}}(t),
\end{cases}
\ee
and
\be
\label{eq4:q_ana_time_evol_sat}
q =
\begin{cases}
    \mathcal{R}_{\mathrm{r}} e^{-\mathcal{K}_{\mathrm{d}} Z}, & Z<Z_{\mathrm{t}}(t), \\
    q_{\mathrm{s}} + \frac{Z_0}{3} \mathcal{K}_{\mathrm{d}} \mathcal{K} \phi^3, & Z_{\mathrm{t}}(t)<Z<Z_{\mathrm{b}}(t),\\
    0, & Z>Z_{\mathrm{b}}(t),
\end{cases}
\ee
% \be
% S=1 \quad \mathrm{for} \quad Z_{\mathrm{t}}(t)<Z<Z_{\mathrm{b}}(t), \ \ t>t_{\mathrm{s}},
% \ee
% and
% \be
% q = q_{\mathrm{s}0} + \frac{Z_0}{3} \mathcal{K}_{\mathrm{d}} \mathcal{K} \phi^3 \quad \mathrm{for} \quad Z_{\mathrm{t}}(t)<Z<Z_{\mathrm{b}}(t), \ \ t>t_{\mathrm{s}},
% \ee
where
\be
\label{eq4:qs0}
q_{\mathrm{s}} = \frac{(3  - Z_0 \mathcal{K}_{\mathrm{d}}) \mathcal{K} \phi_0^3 (Z_{\mathrm{b}}-Z_{\mathrm{t}})}{Z_0 \big( e^{3Z_{\mathrm{b}}/Z_0}-e^{3 Z_{\mathrm{t}}/Z_0} \big)}.
\ee
The unsaturated solution \eqref{eq4:S_ana_time_evol}-\eqref{eq4:q_ana_time_evol} still holds outside of the saturated region.
% We call the boundaries of the saturated region the top and bottom saturation fronts $Z_{\mathrm{t}}(t)$ and $Z_{\mathrm{b}} (t)$, with the lower saturation front is the continuation of the wetting front.
The saturation fronts $Z_{\mathrm{t}}(t)$ and $Z_{\mathrm{b}}(t)$ are found as the numerical solution to two coupled ordinary differential equations (`jump conditions' arising from mass conservation),
% satisfy two coupled ordinary differential equations
\be
\label{eq4:sat_front_ODEs_drainage}
\dot{Z}_{\mathrm{t}} = -\frac{\mathcal{R}_{\mathrm{r}} e^{-\mathcal{K}_{\mathrm{d}} Z_{\mathrm{t}}}-\big(q_{\mathrm{s}} + \frac{Z_0}{3} \mathcal{K}_{\mathrm{d}} \mathcal{K} \phi_{\mathrm{t}}^3 \big)}{\phi_{\mathrm{t}} - (\mathcal{R}_{\mathrm{r}}/\mathcal{K})^{1/3} e^{-\mathcal{K}_{\mathrm{d}} Z_{\mathrm{t}}/3}}, 
\quad \dot{Z}_{\mathrm{b}} = \frac{q_{\mathrm{s}} + \frac{Z_0}{3} \mathcal{K}_{\mathrm{d}} \mathcal{K} \phi_{\mathrm{b}}^3}{\phi_{\mathrm{b}}},
\ee
where $\phi_{\mathrm{t}} = \phi (Z_{\mathrm{t}})$, $\phi_{\mathrm{b}} = \phi (Z_{\mathrm{b}})$,
% We solve the differential equations~\eqref{eq4:sat_front_ODEs_drainage} numerically in MATLAB using \texttt{ode15s}, 
with initial conditions  $Z_{\mathrm{t}} = Z_{\mathrm{b}} = Z_s$ at $t=t_s$.

\subsection{Approximate solution without lateral drainage}

It is of interest to compare solutions with and without lateral drainage.  
The problem without lateral drainage was solved directly in \citet{Meyer2017}, and we can recover that solution from our solutions by taking the limiting case $\mathcal{K}_{\mathrm{d}}\to0$.  
The solutions~\eqref{eq4:S_ana_time_evol} and~\eqref{eq4:q_ana_time_evol} for
% $\tilde{W}$, 
$S$ and $q$ before saturation still hold with $\mathcal{K}_{\mathrm{d}}=0$. Similarly, once saturation is reached, \eqref{eq4:S_ana_time_evol_sat} and~\eqref{eq4:q_ana_time_evol_sat}
still hold with $\mathcal{K}_{\mathrm{d}}=0$, noting that the flux $q=q_{\mathrm{s}}$ is now uniform throughout the saturated region.
% given by
% \be
% \label{eq4:qs}
% q_{\mathrm{s}} = \frac{3 \mathcal{K} \phi_{0}^{3} (Z_{\mathrm{b}} - Z_{\mathrm{t}} )}{Z_0 (e^{3 Z_{\mathrm{b}}/Z_0} - e^{3Z_{\mathrm{t}}/Z_0} ) }.
% \ee

% The behaviour of the wetting and saturation fronts is also different to the drainage case.
The behaviour of the wetting and saturation fronts is recovered from the drainage solutions~\eqref{eq4:Zw_drainage} and~\eqref{eq4:sat_front_ODEs_drainage} by carefully taking the limit $\mathcal{K}_{\mathrm{d}} \to 0$. The wetting front position is
\be
Z_{\mathrm{w}}(t) = \mathcal{R}_{\mathrm{r}}^{2/3} \mathcal{K}^{1/3} t.
\ee
Saturation first occurs at position
\be
\label{eq4:Zs_no_drainage}
Z_{\mathrm{s}} = - \frac{1}{3} Z_0 \log \left( \frac{\mathcal{R}_{\mathrm{r}}}{\mathcal{K} \phi_{0}^{3}} \right),
\ee
and time
\be
\label{eq4:ts_no_drainage}
t_{\mathrm{s}} = - \frac{Z_0}{3 \mathcal{R}_{\mathrm{r}}^{2/3} \mathcal{K}^{1/3}} \log \left( \frac{\mathcal{R}_{\mathrm{r}}}{\mathcal{K} \phi_{0}^{3}} \right).
\ee
Once saturation occurs, the saturation fronts $Z_{\mathrm{t}}(t)$, $Z_{\mathrm{b}}(t)$ satisfy the ordinary differential equations
\be
\label{eq4:sat_front_ODEs}
\dot{Z}_{\mathrm{t}} = \frac{q_{\mathrm{s}} - \mathcal{R}_{\mathrm{r}}}{\phi_{\mathrm{t}} - (\mathcal{R}_{\mathrm{r}}/\mathcal{K})^{1/3}}, \quad \dot{Z}_{\mathrm{b}} = \frac{q_{\mathrm{s}}}{\phi_{\mathrm{b}}},
\ee
with $Z_{\mathrm{t}} = Z_{\mathrm{b}} = Z_s$ at $t=t_s$.
%where $\phi_{\mathrm{t}} = \phi (Z_{\mathrm{t}})$, $\phi_{\mathrm{b}} = \phi (Z_{\mathrm{b}})$ and $q_{\mathrm{s}}$ is given by equation~\eqref{eq4:qs}. 
Once again, we solve the differential equations~\eqref{eq4:sat_front_ODEs} numerically.

\subsection{Influence of lateral drainage}

Figure~\ref{fig4:isothermal_analytical_numerical_evolution} demonstrates the behaviour observed for different values of lateral drainage $\mathcal{K}_{\mathrm{d}}$. Firstly, we see that the wetting front progresses more slowly with lateral drainage than without, as also evident in the logarithmic compared to linear behaviour with time seen in the analytical expressions for $Z_{\mathrm{w}}$. With lateral drainage, there is less water left available to percolate downwards. This also means that saturation occurs at a greater depth (if it ever occurs), and the resulting saturated region grows more slowly.

Moreover, we clearly see the contrasting behaviour depending on whether the lateral drainage strength satisfies $\mathcal{K}_{\mathrm{d}}<3/Z_0$ or $\mathcal{K}_{\mathrm{d}}>3/Z_0$. 
%For small lateral drainage, saturation occurs. 
For large enough lateral drainage, the saturation decreases with depth and saturation never occurs.

In the case when saturation occurs, the depth of the saturation front $Z_{\mathrm{t}}$ tends to a constant value $Z_{\mathrm{t}}^*$ (seen most clearly in the $\mathcal{K}_{\mathrm{d}}=2$ case in figure~\ref{fig4:isothermal_analytical_numerical_evolution}). This can be understood from arguments of mass conservation. In order for the saturated region to grow, there must be more water flowing into it than there is draining out laterally. As the top saturation front gets higher, the lateral drainage $q_{\mathrm{d}}$ out of the saturated region increases, until it approximately balances the downward flux $q$ into the saturated region, i.e. when
\be
q|_{Z=Z_{\mathrm{t}}} \approx \int_{Z_{\mathrm{t}}}^{Z_{\mathrm{b}}} q_{\mathrm{d}} \ \mathrm{d} Z.
\ee
From this balance, we calculate that the limit $Z_{\mathrm{t}}^*$ for the top saturation front $Z_{\mathrm{t}}$ is given by
\be
Z_{\mathrm{t}}^* \approx \frac{1}{3/Z_0 - \mathcal{K}_{\mathrm{d}}} \log \bigg( \frac{Z_0 \mathcal{K}_{\mathrm{d}} \mathcal{K} \phi_0^3}{3 \mathcal{R}_{\mathrm{r}}} \bigg).
\ee
In deriving this expression, we have made use of the fact that the fluxes $q$ and $q_{\mathrm{d}}$ at $Z=Z_{\mathrm{b}}$ are negligible (see figure~\ref{fig4:iso_snapshots_Kd2_10}, $\mathcal{K}_{\mathrm{d}}=2$, $t=1.6,2$), since $\phi$ and hence $q$ and $q_{\mathrm{d}}$ decay exponentially with depth. The analytical expression for $Z_{\mathrm{t}}^*$ agrees with the values obtained from the semi-analytical and numerical solutions, and demonstrates the expected behaviour. The saturated region extends closer to the surface ($Z_{\mathrm{t}}$ decreases) when there is a greater influx $\mathcal{R}_{\mathrm{r}}$ or a smaller porosity $\phi_0$, both of which make it easier to fill the pore space, increasing the extent of the saturated region. Increasing the lateral drainage $\mathcal{K}_{\mathrm{d}}$ has the opposite effect, reducing the extent of the saturated region as more water gets drained away. In particular, we see that the saturated region ends up occupying the whole pore space if $\mathcal{K}_{\mathrm{d}} < 3 \mathcal{R}_{\mathrm{r}}/ ( Z_0 \mathcal{K} \phi_0^3)$.

Note that, despite the arguments above, the lower saturation front $Z_{\mathrm{b}}$ continues to increase over time. We treated the downward flux $q$ at $Z=Z_{\mathrm{b}}$ as negligible when compared to the fluxes nearer the surface, but $\phi$ at $Z=Z_{\mathrm{b}}$ is also very small, enabling the saturated region to continue to slowly grow downwards, despite the very small water flux.
From equation~\eqref{eq4:sat_front_ODEs_drainage} for $\dot{Z}_{\mathrm{b}}$, it can be shown that the large time behaviour of $Z_{\mathrm{b}}$ is given by
\be
Z_{\mathrm{b}} \sim \frac{Z_0}{2} \log (t).
\ee

\section{Steadily melting solutions}
\label{sec4:steadily_melting}

\begin{figure}
    \centering
    \includegraphics[width=\textwidth]{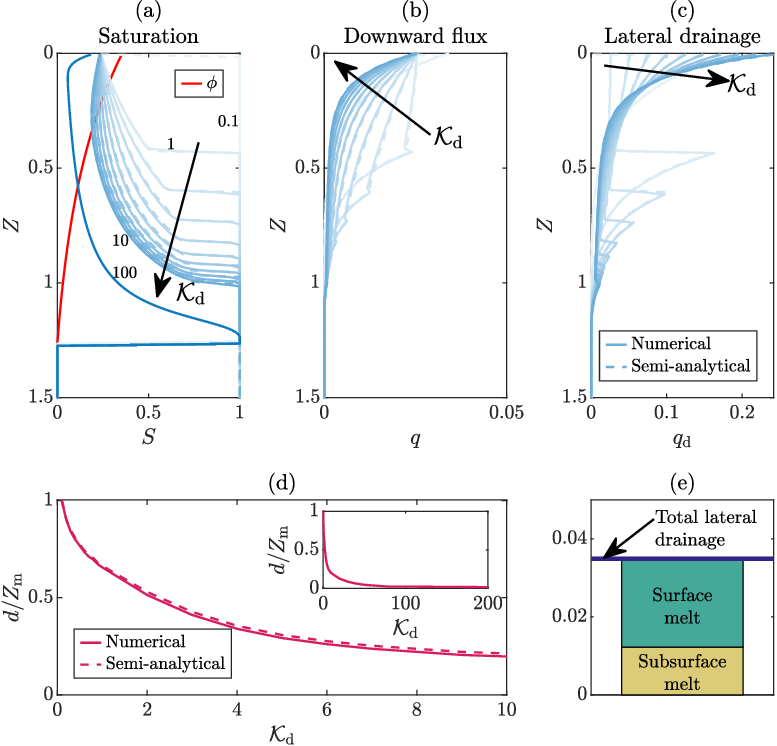}
    % {Figures/steadily_unsat_comb_v2.eps}
    \caption{Steadily melting solutions for different values of the lateral drainage strength $\mathcal{K}_{\mathrm{d}}$. Both the numerical (solid) and semi-analytical (dashed) solutions are shown. Panels (a), (b) and (c) show the profiles of saturation~$S$, downward water flux $q$ and lateral drainage $q_{\mathrm{d}}$, respectively, for different values of $\mathcal{K}_{\mathrm{d}}$. Panel~(d) shows how the proportion $d/Z_{\mathrm{m}}$ of the weathering crust that is saturated with water varies with $\mathcal{K}_{\mathrm{d}}$. Panel~(e) shows the contributions (independent of $\mathcal{K}_{\mathrm{d}}$) to the total lateral drainage from surface melt and the subsurface melt when the surface is unsaturated (see equation \eqref{eq4:total_lateral_drainage}). 
    %Note that semi-analytical solutions are not shown for $\mathcal{K}_{\mathrm{d}}=100$ in panel~(a), (b), (c) or in the inset of panel~(d). 
    Parameter values: $\mathcal{B} = 10^4$, $\mathcal{Q}_{\mathrm{si}} = 1$, $\mathcal{Q}_0 = 0.25$, with the other parameter values as shown in table~\ref{tab4:dimensionless_parameters}. Solutions shown are dimensionless.}
    \label{fig4:steadily_unsat_comb}
\end{figure}

Next, we seek `steadily melting' solutions. These are solutions to the model with constant forcings $\mathcal{Q}_{\mathrm{si}}$ and $\mathcal{Q}_0$ such that the surface lowers at a constant rate $\dot{h} < 0$, the weathering crust occupies a region between the surface and a constant depth $Z_{\mathrm{m}}$, and the temperature, porosity and saturation profiles in the $Z$-frame are constant in time (i.e. the situation in figure~\ref{fig4:WC_water_table_schematic_3D_1D}(b), independent of time). In \citet{Woods2023}, solutions were found for the case when there is no flow and the crust is fully saturated. Here, we seek steadily melting solutions for the problem with lateral drainage, which also therefore involve vertical percolation and a water table.

The ice mass conservation equation, energy conservation equations and their boundary conditions under the steadily melting assumptions are independent of saturation. Hence the steadily melting solutions for $T$, $\phi$, $\dot{h}$, and $Z_{\mathrm{m}}$ are identical to the solutions for the saturated (no-drainage) case presented in \citet{Woods2023}. The dimensionless forms of these solutions are
\begin{equation}
\label{eq4:T_sm_nondim}
T = 
\begin{cases} 
0, & 0<Z<Z_{\mathrm{m}}, \\
\begin{split}
- \frac{(1-\chi)(1-a) \mathcal{Q}_{\mathrm{si}}}{\hat{k} \hat{\kappa} + \dot{h}} \bigg( e^{-\hat{\kappa} Z} - e^{-\hat{\kappa} Z_{\mathrm{m}} } e^{\frac{ \dot{h}}{\hat{k}} \big( Z - Z_{\mathrm{m}} \big)} \bigg)\\
+ \mathcal{T}_{\infty} \bigg( 1 - e^{\frac{\dot{h}}{\hat{k}} \big( Z - Z_{\mathrm{m}} \big)} \bigg)
\end{split},& Z>Z_{\mathrm{m}},\\
\end{cases}
\end{equation}
\begin{equation}
\label{eq4:phi_sm_nondim}
\phi = 
\begin{cases} 
\begin{aligned} 
- \frac{(1-\chi)(1-a) \mathcal{Q}_{\mathrm{si}}}{\dot{h} \mathcal{S}} \bigg( e^{-\hat{\kappa} Z} - e^{-\hat{\kappa} Z_{\mathrm{m}}} \bigg),
\end{aligned}
& 0<Z<Z_{\mathrm{m}}, \\
0, & Z>Z_{\mathrm{m}}, \\
\end{cases}
\end{equation}
\begin{equation}
\label{eq4:hdot_sm_nondim}
\dot{h} = -\frac{(1-a) \mathcal{Q}_{\mathrm{si}} + \mathcal{Q}_0}{ \mathcal{S}  - \mathcal{T}_{\infty}},
\end{equation}
\begin{equation}
\label{eq4:Zm_sm_nondim}
Z_{\mathrm{m}} = \frac{1}{\hat{\kappa}} \log \Bigg( \bigg(\frac{(1-a)(1-\chi) \mathcal{Q}_{\mathrm{si}}}{ - \mathcal{T}_{\infty} } \bigg)  \bigg/  \bigg( \frac{(1-a) \mathcal{Q}_{\mathrm{si}} + \mathcal{Q}_0}{\mathcal{S}  - \mathcal{T}_{\infty}} \bigg) \Bigg).
\end{equation}
The only difference from the saturated solution is that now some of the porous region is unsaturated. We expect that the solution will have two layers: a saturated layer immediately above the impermeable base, with the rest of the weathering crust above this being unsaturated, as shown in figure~\ref{fig4:WC_water_table_schematic_3D_1D}. We refer to the boundary between the saturated and unsaturated regions as the water table, $Z=Z_{\mathrm{d}}$, and write $d = Z_{\mathrm{m}}-Z_{\mathrm{d}}$ for the thickness of the saturated layer.

\subsection{Approximate solution}

As for the isothermal water transport example, we neglect capillary pressure (assuming $\mathcal{B} \gg 1$) to enable us to find a semi-analytical solution. We seek solutions for the saturation $S$ and the water table position $Z\sub{d}$.

Writing $W = 1-\phi+\phi S$ for the total water content, the steadily melting form of the mass conservation equations~\eqref{eq4:mass_ice_unsat_nondim}-~\eqref{eq4:mass_water_unsat_nondim} is
\be
\label{eq4:steadily_Weqn}
\dot{h} \dd{W}{Z} + \dd{q}{Z} = - q_{\mathrm{d}} \quad \mathrm{with} \quad q_{\mathrm{d}} = \mathcal{K}_{\mathrm{d}} \mathcal{K} \phi^3 S^3,
\ee
with boundary conditions
\be
\label{eq4:steadily_W_surface_BC}
q + \dot{h} W = 0 \quad \mathrm{at} \quad Z=0,
\ee
\be
\label{eq4:steadily_W_Zm_BC}
q = 0 \quad \mathrm{at} \quad Z=Z_{\mathrm{m}}.
\ee
Here, we are assuming that the surface is unsaturated ($S<1$) in~\eqref{eq4:steadily_W_surface_BC}, and hence that the surface runoff $r_{\mathrm{off}}$ is zero.

% To find our solution, we need to solve for the saturation $S$ in the unsaturated region, the flux $q$ in the saturated 

% We need to determine the saturation $S$ in the unsaturated region and the position of the water table $Z_{\mathrm{d}}$. To do this,

Our method of solution is to solve equation \eqref{eq4:steadily_Weqn} for the water flux $q$ in the saturated region using the boundary condition~\eqref{eq4:steadily_W_Zm_BC} at $Z=Z_{\mathrm{m}}$, and solve~\eqref{eq4:steadily_Weqn} for the saturation $S$ in the unsaturated region using the boundary condition~\eqref{eq4:steadily_W_surface_BC} at $Z=0$ (details given below). We then determine $Z_{\mathrm{d}}$ as the point at which $q+\dot{h}W$ for these two solutions are equal. (Integrating~\eqref{eq4:steadily_Weqn} across the water table shows that $q+\dot{h} W$ must be continuous there.)

In the saturated region, $W=1$, and so equation~\eqref{eq4:steadily_Weqn} is
\be
\label{eq4:steadily_q_sat_eqn}
\dd{q}{Z} = -q_{\mathrm{d}} \quad \mathrm{with} \quad q_{\mathrm{d}} = \mathcal{K}_{\mathrm{d}} \mathcal{K} \phi^3,
\ee
which is solved with boundary condition~\eqref{eq4:steadily_W_surface_BC} and the steadily melting form of the porosity~\eqref{eq4:phi_sm_nondim} to give
\be
\begin{split}
\label{eq4:q_steadily_melting_sat}
q = -\mathcal{K}_{\mathrm{d}} \mathcal{K} \phi_{\mathrm{sm0}}^3 \bigg( -\frac{1}{3\hat{\kappa}} \big( e^{-3 \hat{\kappa} Z} - e^{-3 \hat{\kappa} Z_{\mathrm{m}}} \big) + \frac{3}{2 \hat{\kappa}} \big( e^{-2 \hat{\kappa} Z} e^{-\hat{\kappa} Z_{\mathrm{m}}} - e^{-3 \hat{\kappa} Z_{\mathrm{m}}} \big) \\
- \frac{3}{\hat{\kappa}} \big( e^{-\hat{\kappa} Z} e^{-2 \hat{\kappa} Z_{\mathrm{m}}} - e^{-3 \hat{\kappa} Z_{\mathrm{m}}} \big) - e^{-3 \hat{\kappa} Z_{\mathrm{m}}} (Z-Z_{\mathrm{m}}) \bigg) ,
\end{split}
\ee
where
\be
\phi_{\mathrm{sm0}} = - \frac{(1-\chi)(1-a) \mathcal{Q}_{\mathrm{si}}}{\dot{h} \mathcal{S}}.
\ee

Under the assumption that $\mathcal{B} \gg 1$, the downward water flux in the unsaturated region is $q = \mathcal{K} \phi^3 S^3$, so equation~\eqref{eq4:steadily_Weqn} becomes
\be
\label{eq4:steadily_unsat_region_eqn}
\dd{}{Z} \Big( \mathcal{K} \phi^3 S^3 + \dot{h} (1 - \phi + \phi S) \Big) = -\mathcal{K}_{\mathrm{d}} \mathcal{K} \phi^3 S^3,
\ee
with the boundary condition~\eqref{eq4:steadily_W_surface_BC} becoming
\be
\label{eq4:steadily_unsat_region_BC}
\mathcal{K} \phi^3 S^3 + \dot{h} (1 - \phi + \phi S) = 0 \quad \mathrm{at} \quad Z=0.
\ee
This boundary condition uniquely determines the value of $S$ 
%(and hence $q = \mathcal{K} \phi^3 S^3$) 
at the surface, independent of $\mathcal{K}_{\mathrm{d}}$ (see figure~\ref{fig4:steadily_unsat_comb}). We solve this unsaturated problem for $S$ numerically in MATLAB using \texttt{fsolve} for the boundary condition~\eqref{eq4:steadily_unsat_region_BC} and \texttt{ode15s} for the differential equation~\eqref{eq4:steadily_unsat_region_eqn}. We then calculate $q = \mathcal{K} \phi^3 S^3$ and use continuity of $q+\dot{h}W$ at $Z=Z_{\mathrm{d}}$ to determine $Z_{\mathrm{d}}$.

\subsection{Influence of lateral drainage}

Figure~\ref{fig4:steadily_unsat_comb} shows numerical and semi-analytical solutions to the steadily melting problem for different values of lateral drainage strength $\mathcal{K}\sub{d}$. The two sets of solutions agree well. The numerical solutions use $\mathcal{B} = 10^4$, so the large Bond number assumption made in the semi-analytical solution is reasonable.

In the solution with $\mathcal{B} \gg 1$, the saturation $S$ is discontinuous at $Z=Z\sub{d}$. Therefore, the condition that $q+\dot{h}W$ is continuous at $Z=Z_{\mathrm{d}}$ means that $q$ must also be discontinuous at $Z=Z\sub{d}$. Since $\dot{h}<0$, $q$ must be larger on the saturated side, as we see in figure~\ref{fig4:steadily_unsat_comb}(b). Similarly, since $q\sub{d} \sim S^3$, there is also a jump increase in the lateral drainage flux as we pass from unsaturated to saturated across $Z=Z\sub{d}$.

The thickness $d$ of the saturated region decreases with increasing lateral drainage strength $\mathcal{K}\sub{d}$ (see figure~\ref{fig4:steadily_unsat_comb}(a,d)). Despite this, the total lateral drainage $\int q\sub{d} \; \mathrm{d} Z$ is independent of $\mathcal{K}\sub{d}$, as long as the surface remains unsaturated. The difference is that, with larger $\mathcal{K}\sub{d}$, more water is drained away nearer the surface, so less water is available to percolate downwards to the saturated layer. This is reflected in the flux profiles in figure~\ref{fig4:steadily_unsat_comb}(b,c).

The reason why the total lateral drainage is independent of $\mathcal{K}\sub{d}$ is due to mass conservation.  This can be seen by integrating the steadily melting mass conservation equation \eqref{eq4:mass_water_unsat_nondim},
\be
\dot{h} \pd{}{Z} \big( \phi S \big) + \pd{q}{Z} = m_{\mathrm{int}} - q_{\mathrm{d}},
\ee
over the weathering crust ($Z=0$ to $Z=Z_{\mathrm{m}}$), with the no flux condition at $Z=Z_{\mathrm{m}}$ and the kinematic condition~\eqref{eq4:kinematic_water_unsat} applied at the surface (neglecting rainfall here).
% This results in the total lateral drainage $\int q_{\mathrm{d}} \ \mathrm{d} Z $ being a sum of the total subsurface melt $\int m_{\mathrm{int}} \ \mathrm{d} Z$ and the surface melt $m_{\mathrm{surf}}$,
This gives
\be
\label{eq4:total_lateral_drainage}
\int q_{\mathrm{d}} \ \mathrm{d} Z + r_{\mathrm{off}} = \int m_{\mathrm{int}} \ \mathrm{d} Z + m_{\mathrm{surf}} ,
\ee
which says that the total runoff rate (subsurface lateral drainage + surface runoff) must match the total melt rate (subsurface + surface). If $\mathcal{K}_{\mathrm{d}}$ is sufficiently large that the surface is unsaturated then $r_{\mathrm{off}}$ is zero and the total lateral drainage $\int q_{\mathrm{d}} \ \mathrm{d} Z$ must simply balance the subsurface and surface melt sources, as shown in figure~\ref{fig4:steadily_unsat_comb}(e).

% If the surface is saturated (i.e. if $\mathcal{K}_{\mathrm{d}}$ is less than about 0.1), it is the sum of the total lateral drainage and the surface runoff that is independent of $\mathcal{K}_{\mathrm{d}}$.

\subsection{Inferring the value of $\mathcal{K}_{\mathrm{d}}$}
% \begin{itemize}
%     \item Could move into discussion? e.g. When talking about how the model could be applied in real life.
% \end{itemize}

% \tw{Could move into discussion? e.g. When talking about how the model could be applied in real life, or what the idealised solutions can tell us.}

% Our model results show that the thickness of the saturated region decreases with increasing lateral drainage strength $\mathcal{K}_{\mathrm{d}}$ (figure~\ref{fig4:steadily_unsat_comb}(d)). 
The decreasing relationship between saturated layer thickness $d$ and lateral drainage strength $\mathcal{K}_{\mathrm{d}}$, shown in figure~\ref{fig4:steadily_unsat_comb}(d), can help us to infer a suitable value for $\mathcal{K}_{\mathrm{d}}$ based on observations of the water table height. Typically, most of the weathering crust is saturated with meltwater. For example, in Greenland, the water table has been observed typically to lie $\sim 0.1$ m below the surface in weathering crusts that are $\sim 1$ m thick \citep{Cooper2018,Irvine-Fynn2021}. Therefore, we expect $d/Z_{\mathrm{m}}$ closer to 1 than~0, suggesting the appropriate order of magnitude is $\mathcal{K}_{\mathrm{d}} \sim 1$. However, the transient behaviour of the weathering crust and meltwater flow means that a different value of $\mathcal{K}_{\mathrm{d}}$ might prove to be more appropriate in a transient setting.

An alternative method for inferring the value of $\mathcal{K}_{\mathrm{d}}$ is to recall the definition in \eqref{eq4:weathering_crust_dimensionless_parameters} and use observations of the surface slope $\sin\theta$ and an appropriate lengthscale $L_{\mathrm{d}}$ over which lateral drainage occurs, such as the distance between supraglacial streams. For example, at a study site in south-western Greenland, \citet{Irvine-Fynn2021} measured the surface slope ($\sin\theta$) to be 0.022 m m$^{-1}$ and the mean distance to a surface stream ($L_{\mathrm{d}}$) to be 7.24 m, giving $\sin\theta /L_{\mathrm{d}} \approx 0.003$. However, this method would also require an estimation of the anisotropy factor $\beta_{\mathrm{h}}$ to calculate $\mathcal{K}_{\mathrm{d}}$.

\section{Periodic switching}
\label{sec:periodic_switching}
\begin{figure}
    \centering
    \includegraphics[width=0.9\linewidth]{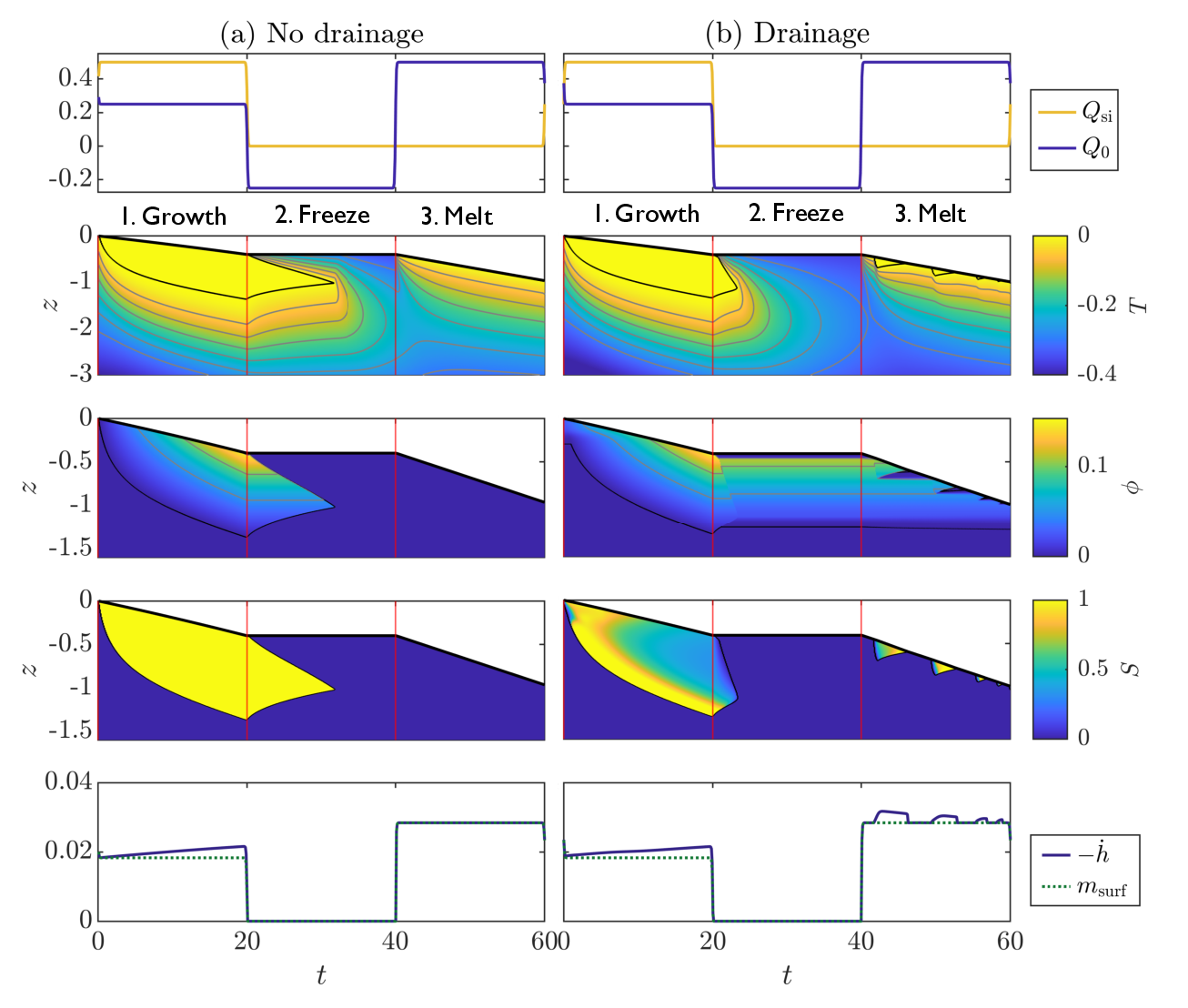}
    % {Figures/periodic_switch_plot_comb_Kd20.eps}
    \caption{Evolution of the temperature $T$, porosity $\phi$, saturation $S$, surface lowering rate $-\dot{h}$ and surface melt rate $m_{\mathrm{surf}}$ during the periodic switching experiment (a) without and (b) with lateral drainage ($\mathcal{K}_{\mathrm{d}}=0$ and $\mathcal{K}_{\mathrm{d}}=20$, respectively). Black lines in the $T$, $\phi$ and $S$ plots show the zero contours of the respective variables. Grey lines in in the $T$ and $\phi$ plots both show contours of spacing 0.05. Red vertical lines show the transitions between stages 1, 2 and 3. The radiative forcings are shown in the top panels.}
    \label{fig:periodic_switch}
\end{figure}

% FROM IMA PAPER!
% The weathering crust is a naturally transient structure. It only forms in the summer months when there is enough shortwave radiation to cause subsurface (subsurface) melting \citep{Cook2016}. On a shorter timescale, the weathering crust responds to rapidly changing weather conditions, growing and decaying on the order of hours and days \citep[e.g.][]{Schuster2001,Tedstone2020}.
% Clear, sunny conditions favour formation of the crust, whereas warm, windy, cloudy conditions favour removal \citep{Muller1969,Schuster2001}. 

Weathering crusts evolve on short timescales in response to rapidly changing weather conditions \citep{Schuster2001,Tedstone2020}, growing in clear, sunny conditions and being removed in warm, windy, cloudy conditions \citep{Muller1969}. A weathering crust also requires sufficiently high levels of shortwave radiation to form (i.e. during the summer) \citep{Cook2016}. We now consider an idealised `periodic switching' experiment, with piecewise-constant forcings, that captures three key stages of weathering crust evolution. Within each of the three stages of the experiment, the forcings are held constant (as shown in figure~\ref{fig:periodic_switch}), and the three stages are repeated periodically. We refer to these stages using the following terms:
\begin{enumerate}
    \item \emph{Growth}. Growing a weathering crust via shortwave-radiation-induced subsurface melting. ($\mathcal{Q}\sub{si}>0$ and sufficiently large).

    \item \emph{Freeze}. Freezing the liquid water within the weathering crust. ($\mathcal{Q}_0<0,\mathcal{Q}\sub{si}=0$).

    \item \emph{Melt}. Melting from the surface. ($\mathcal{Q}_0>0,\mathcal{Q}\sub{si}=0$).
\end{enumerate}
The freeze and melt stages both represent mechanisms by which a weathering crust can be removed, and were previously examined in \citet{Woods2024} in the absence of lateral drainage. 

The results of the periodic switching experiment are shown in figure~\ref{fig:periodic_switch}, where we compare solutions with and without lateral drainage. As in our previous examples, the weathering crust is always saturated with water in the no-drainage case but can be unsaturated in the drainage case. 
We will now discuss the key processes demonstrated by these simulations and the differences between the drainage and no-drainage solutions, stage-by-stage.

\subsection{Growth}

Growth of the weathering crust is driven by subsurface shortwave radiation causing subsurface melting. The growth behaviour is very similar in both no-drainage and drainage cases. This is because the subsurface heat source $-\pd{F}{Z} = \hat{\kappa}(1-\chi)(1-a) \mathcal{Q}\sub{si}$, which drives subsurface melting and temperature evolution, is independent of saturation $S$. (Note that this is a consequence of assuming that the radiation is independent of saturation; the potential impact of saturation on absorption and scattering coefficients could be incorporated with a more involved two-stream approximation for $F$, similar to what was done in \citet{Woods2023} to investigate the effect of microbes in the water.)

% \color{blue}
% ALTERNATIVE!!!
% This is explained by the subsurface heat source $-\pd{F}{Z} = \hat{\kappa}(1-\chi)(1-a) \mathcal{Q}\sub{si}$ being independent of saturation in the present manuscript. This means that the temperature and porosity evolution are independent of saturation, provided that no freezing occurs (we will discuss the freezing case below). We see this independence from the mass and energy conservation equations \eqref{eq4:mass_ice_unsat_nondim}, \eqref{eq4:temp_eqn_unsat_nondim}. Crucially, we note that the latter is an equation for the temperature in cold regions (where $S$ is no well-defined, so set to zero), and an equation for the subsurface melt rate $m\sub{int}$, namely
% \be
% m\sub{int} = - \frac{1}{\mathcal{S}} \pd{F}{Z},
% \ee
% in temperate regions. In both cases, there is no explicit $S$-dependence, and nor is there in equation \eqref{eq4:mass_ice_unsat_nondim} for $\phi$. \tw{Put some of this detail in an appendix, or just say less in the first place? Or discuss more generall in discussion (when mentioning that making absorption/scattering depend on $S$ (or $\phi$) could change things).}

% \color{black}

% As it happens, the lateral drainage is weak enough compared to the source of water from melting that the crust remains saturated during most of the growth phase (once the initial porous surface layer has `filled up').

% However, the weathering crust is saturated only in a thin layer at the bottom in the drainage case, compared to being completely saturated in the no drainage case.

However, the saturation profiles differ between the two cases. In the drainage case, once the crust has `filled up' at the start of the growth phase, the crust is fully saturated for only a short time until the porosity - and hence lateral drainage flux - becomes large enough that subsurface meltwater removal dominates melt. After this, the crust is mostly unsaturated with a thin saturated layer at the bottom, in contrast to the fully-saturated no drainage case. Note that with smaller values of $\mathcal{K}\sub{d}$ (e.g. $\mathcal{K}\sub{d}=3$, not shown), this transition never occurs and the crust remains mostly saturated throughout, consistent with the behaviour in section~\ref{sec4:isothermal_water}.

% \textcolor{red}{Update.} In fact, the weathering crust is saturated throughout most of the growth stage even in the drainage case, because the lateral drainage flux is not larger than the water supplied by (surface) melting. There is a small unsaturated region at the start of the growth phase due to there initially being a small, dry, porous surface layer, which takes a little time to fill up with meltwater.

\subsection{Freeze}

\begin{figure}
    \centering
    \includegraphics[width=\textwidth]{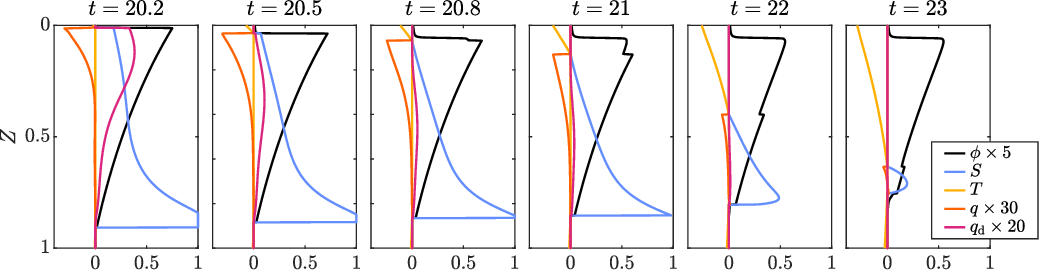}
    % {Figures/periodic_switch_snapshots_Kd20_06-Oct-2025.eps}
    \caption{Snapshots of the freezing stage (stage 2) of the periodic switching experiment with drainage ($\mathcal{K}_{\mathrm{d}}=20$) shown in figure~\ref{fig:periodic_switch}. The porosity $\phi$ (black), saturation $S$ (blue), temperature $T$ (yellow), downward water flux $q$ (orange), and the lateral drainage flux $q_{\mathrm{d}}$ (pink) are shown. Note that scaled versions of the variables are plotted to make them more visible, and that the time spacing is non-uniform.}
    \label{fig:periodic_switching_snapshots}
\end{figure}

During the freezing stage of our experiment (figure~\ref{fig:periodic_switch}, stage 2), all the water in the weathering crust is frozen, by freezing fronts that propagate from both above and below, and the ice subsequently cools. This is achieved by having an absence of shortwave radiation ($\mathcal{Q}\sub{si}=0$) and hence subsurface heating, and a negative surface forcing ($\mathcal{Q}_0<0$). We now discuss the key differences between the drainage and no-drainage cases during the freezing stage.

Freezing and cooling happens more quickly in the drainage case because the weathering crust is unsaturated.
Freezing releases latent heat, so when there is less water to freeze, the ice can cool more quickly. The freezing front also moves more quickly.  This can be seen from the Stefan condition, which must hold at the freezing interfaces:
\be
\label{eq:Stefan}
\bigg[ -\hat{k} \pd{T}{Z} \bigg]_{-}^{+} = \dot{Z}\sub{f} \mathcal{S} [\phi S]_{-}^{+} \quad \mathrm{at} \quad Z=Z\sub{f},
\ee
where $Z\sub{f}$ is the position of a freezing front, and $[ \cdot ]_{-}^{+}$ represents a jump across the freezing front. 
%On the temperate (i.e. not-yet-frozen) side of the freezing front, $\pd{T}{Z}=0$.

Another key difference is that the frozen ice formed in the drainage case is often porous, as compared to the solid ice that forms in the no-drainage case. 
% We also observe the effect of capillary pressure in the drainage case. It produces a layer of near-solid ice at the surface. 

However, capillary pressure effects generally cause a layer of near-solid ice to form near the surface, even in the drainage case (see figure~\ref{fig:periodic_switching_snapshots}).
When the freezing phase begins, the vertical water flux $q$ becomes negative (upward flow) as a result of the capillary pressure term in the flux~\eqref{eq4:Darcy_nondim} driving the flow down the saturation gradient, towards the surface (figure~\ref{fig:periodic_switching_snapshots}, $t=20.2$). This provides a sufficient supply of water to the upper freezing front, where the water freezes, producing near-solid ice (figure~\ref{fig:periodic_switching_snapshots}, $t=20.2$ to 20.4). This continues until the saturation drops to zero at the freezing front. At this point, there is no longer a supply of water to fill the pore space and freeze. Instead, the freezing front continues to propagate with only a small amount of water being frozen there, resulting in the frozen ice being much more porous (figure~\ref{fig:periodic_switching_snapshots}, $t=20.8$ to 23).

\subsection{Melt}

During the melt stage, melting occurs at the surface only (not internally). This is achieved by having positive surface forcing ($\mathcal{Q}_0>0$) and no shortwave radiation ($\mathcal{Q}\sub{si}=0$). Melt at the surface results in the surface lowering at a rate $-\dot{h}$, given by the kinematic condition~\eqref{eq4:kinematic_ice_nondim}.

Lateral drainage can have two opposing effects on the rate of surface lowering.
On one hand, it can result in the surface ice being porous (rather than solid), which can enhance surface lowering.
A given amount of melt energy determines the \emph{volume} of ice that is melted, and for porous ice, this corresponds to a greater \emph{thickness} of ice.
On the other hand, the ice in the drainage case is colder (both in the porous layer and in the solid ice below), so more energy is required to heat up the ice to the melting temperature before it can be melted. 

In our simulations, we observe that the former effect dominates: surface lowering is very slightly enhanced (by 2\%) by lateral drainage. We can see this enhancement in the small bumps in $-\dot{h}$ in figure~\ref{fig:periodic_switch}(b), which occur when porous ice is being melted at the surface.
% \ijh{where do we see this?  Is this just the tiny blip in $\dot{h}$ in figure 5(b)?  It is only a very slight difference in that case?}

\begin{figure}
    \centering
    \includegraphics[width=\textwidth]{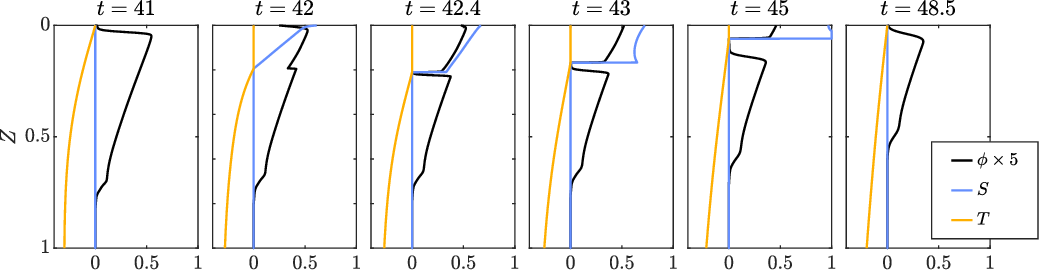}
    % {Figures/periodic_switch_snapshots_Kd20_melt_06-Oct-2025.eps}
    \caption{Snapshots of the formation of an impermeable ice layer during the melt stage (stage 3) of the periodic switching experiment with drainage ($\mathcal{K}\sub{d}=20$) shown in figure~\ref{fig:periodic_switch}. The porosity $\phi$ (black), saturation $S$ (blue) and temperature $T$ (yellow) are shown. Note that a scaled version of $\phi$ is plotted to make it more visible, and that the time spacing is non-uniform.}
    \label{fig:periodic_switch_snapshots_melt}
\end{figure}

Despite the weathering crust being porous in the drainage case, very little meltwater infiltrates during the melt phase. This is due to thin, impermeable layers forming at the wetting front (at the bottom of the near-surface meltwater `blips' in figure~\ref{fig:periodic_switch}(b), shown in detail in figure~\ref{fig:periodic_switch_snapshots_melt}). As the meltwater infiltrates into cold ice ($T<0$), some water freezes at the wetting front (figure~\ref{fig:periodic_switch_snapshots_melt}, $t=42$). At a certain depth, the porosity is small enough that the freezing water fills the entire pore space, creating an impermeable layer (figure~\ref{fig:periodic_switch_snapshots_melt}, $t=42.4$). The layer persists (and in fact grows) until it is melted through via surface melting (figure~\ref{fig:periodic_switch_snapshots_melt}, $t=48.5$). For some time after that, the surface porosity is low enough that the surface instantly becomes impermeable due to freezing surface meltwater, preventing any infiltration. We see this entire process occurring four times in figure~\ref{fig:periodic_switch}(b).

\citet{Shadab2024} also observe the formation of very-low-porosity layers (`ice lenses') in their glacial-firn model, but via a different mechanism: ice lenses form in homogeneous firn (with uniform porosity) after meltwater pulses, when the drying front (at the end of the meltwater pulse) catches up with the wetting/freezing front. In both models, ice lenses form in the absence pre-existing discontinuities (e.g. impurities).
%We note that the two models are almost identical, aside from our addition of subsurface shortwave radiation, subsurface lateral drainage, and capillary forces in Darcy's law (as well as a different choice of exponent in the relative permeability function).

% It is interesting to compare the formation of impermeable layers in our model to the formation of ice lenses (impermeable layers) in the glacial-firn simulations by \citet{Shadab2024}. We note that the two models are almost identical, aside from our addition of subsurface shortwave radiation, subsurface lateral drainage, and capillary forces in Darcy's law (as well as a different choice of exponent in the relative permeability function). \citet{Shadab2024} demonstrated that it is possible to form ice lenses in homogeneous, cold firn after a pulse of meltwater input, when the drying front (at the end of the meltwater pulse) catches up with the wetting/freezing front. Here, we see that --- provided porosity decreases with depth --- it is possible for ice lenses to form whilst there is continued meltwater input.

\subsection{Conservation of water}
\label{sec:water_conservation_periodic}

\begin{figure}
    \centering
    \includegraphics[width=0.95\linewidth]{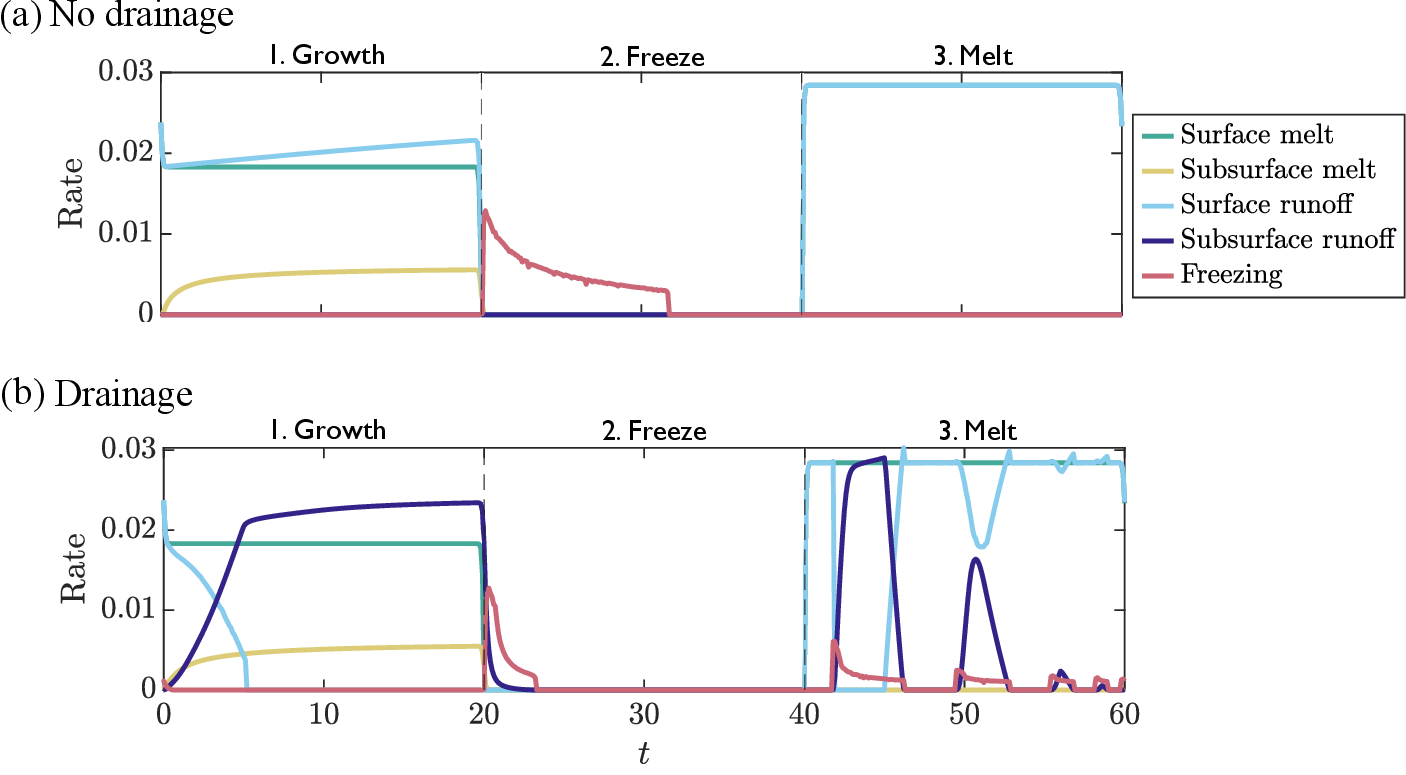}
    \caption{Evolution of the rates at which meltwater is produced (via surface melting and subsurface melting) and removed (via surface runoff, subsurface runoff and subsurface freezing) over one period of the periodic switching experiment shown in Figure~\ref{fig:periodic_switch}, (a) without and (b) with lateral drainage ($\mathcal{K}\sub{d}=0$ and $\mathcal{K}\sub{d}=20$, respectively).} 
    \label{fig:melt_runoff_evol}
\end{figure}

Meltwater is produced by surface melting and subsurface melting, and can be removed by refreezing, surface runoff or subsurface runoff (lateral drainage). Figure~\ref{fig:melt_runoff_evol} shows the evolution of these 5 components over one period for the cases with and without lateral drainage. We plot the rates of surface melt $m\sub{surf}$, subsurface melt $\int \max (m\sub{int},0) \ \mathrm{d} Z$, subsurface freezing $-\int \min (m\sub{int},0) \ \mathrm{d} Z$, surface runoff $r\sub{off}$ and subsurface runoff $\int q\sub{d} \ \mathrm{d}Z$. (Note that the subsurface melt rate $m_{\mathrm{int}}$ corresponds to melting when positive and freezing when negative.) During the melt phase, melting occurs purely at the surface. In the no-drainage case, all of this surface meltwater runs off at the surface. In contrast, in the drainage case, some surface meltwater penetrates and is instead removed by subsurface runoff or freezing. There are small `blips' during which the surface runoff rate exceeds the surface melt rate. This is due to water already held in the porous, saturated ice also being removed as surface runoff when the porous ice melts from the surface. 

\section{Summary and discussion}
\label{sec:conclusion}

Radiation-driven melt and meltwater flow are two important processes affecting near-surface heating, melting and runoff at ice surfaces. Here, we have captured both of these processes in a single model by including an energy equation that includes radiative energy fluxes, meltwater flow driven by Darcy's law, and a Darcy-based parametrisation to account for lateral flow. This allows the weathering crust to become unsaturated.

Our model was applied to idealised scenarios, to explore the effect of meltwater flow and lateral drainage on the weathering crust. We solved the model equations numerically using an enthalpy method. We also considered two test cases for which we could find semi-analytical solutions in the limit of small capillary pressure ($\mathcal{B}\gg1$). This allowed us to validate our numerical solutions.

The model with lateral drainage produces a weathering crust that has a saturated layer at the base, above which it is unsaturated, as is observed in the field \citep[e.g.][]{Irvine-Fynn2021}. 
We observe that a greater lateral drainage strength $\mathcal{K}_{\mathrm{d}}$ results in a thinner saturated layer (a lower water table). This relationship potentially provides a method for inferring an appropriate value of $\mathcal{K}_{\mathrm{d}}$ from observations of water table height.

Furthermore, we have observed that including lateral drainage alters the porosity profile of the weathering crust and can enhance surface lowering, 
by allowing the weathering crust to refreeze with non-zero porosity, making it easier to subsequently re-melt than solid ice. 
This contrasts with the findings of \citet{Hoffman2014} that the inclusion of lateral drainage in their one-dimensional model led to a \emph{decrease} in the amount of surface lowering. However, this is explained by the fact that \citet{Hoffman2014} also used different values of the shortwave radiation partitioning parameter $\chi$ for their models with and without lateral drainage, and the value of $\chi$ also has a significant effect on surface lowering \citep[e.g.][]{Hoffman2014,Law2020}.

Another interesting result from the lateral drainage model is the formation of a transient, impermeable ice layer, analogous to the `ice lenses' formed in the glacial-firn model of \citet{Shadab2024}.
% (during the `melt' stage of figure~\ref{fig4:drainage_switching_evol}(b)). 
It is observed that weathering crusts can contain alternating layers of porous and impermeable ice \citep[e.g.][]{Cooper2018}.
As seen in the `melt' stage of figure~\ref{fig:periodic_switch}(b), the formation of impermeable layers can prevent the infiltration of meltwater into the porous ice below, potentially enhancing lateral meltwater flow and runoff, as also observed in snow and firn \citep[e.g.][]{Mikkelsen2016,Noel2017,Clerx2022}.

However, there are some aspects of observed weathering crust evolution that our model does not yet fully capture. In particular, \citet{Stevens2026} observed that the density of the weathering crust increases at all depths simultaneously during weathering crust removal, in contrast to our model results in which the density increases only at the freezing front (e.g. figure~\ref{fig:periodic_switching_snapshots}). One possible explanation for this is our assumption of local thermodynamic equilibrium, which could be relaxed by using a non-equilibrium thermodynamics model \citep[e.g.][]{Moure2023}.
%but at the cost of significantly increasing the model complexity.

% There are a few ways in which our model could be further developed.
There are a number of further approximations and assumptions made in this study which may need to be relaxed in future work.
We assumed that the optical properties of the weathering crust are constant and uniform. In reality, the absorption and scattering of subsurface shortwave radiation will depend on the structure and water content of the ice ($\phi$ and $S$), which can lead to a potential feedback where the presence of more meltwater lowers the albedo and enhances further melting. This behaviour could be incorporated using the two-stream approximation presented in \citet{Woods2023}. 
% Including this in the model was briefly considered in \citet{Woods2023}, using the two-stream approximation.

% Our model provides a starting point for exploring various aspects of weathering crust behaviour. For example, how the weathering crust affects the transport of meltwater across the ice sheet surface. 

An obvious avenue for future work is to extend the model to higher dimensions, with a more explicit treatment of the lateral drainage. This would allow us to study the spatial-temporal variability of the weathering crust and the associated albedo, as well as the transport of meltwater through the weathering crust across the ice surface. A full two- or three-dimensional model \citep[e.g.][for snow]{Leroux2017, Moure2023} would be insightful but computationally costly. An alternative approach to capture the key behaviour is to derive a depth-integrated form of our model, making use of the small aspect ratio of the weathering crust. This method would be analogous to the Dupuit approximation commonly used for modelling groundwater flow \citep{Fowler2011}. A similar, simplified approach (although not depth-integrated) has previously been used by \citet{Miller2020,Miller2022} to model firn aquifers.

Although our focus here was idealised scenarios,
the model can also be applied to real-world scenarios, such as locations on the Greenland Ice Sheet.
The model forcings $Q\sub{si}$, $Q\sub0$ and $\upsilon$ can be calculated 
% \citep[e.g. using the method in][]{Woods2023} 
from observed (or modelled) meteorological conditions 
% , which are readily available across the Greenland Ice Sheet 
\citep[e.g.][]{Fausto2021}. So far, \citet{Stevens2026} have applied the no-drainage version of the model \citep{Woods2024} to the Greenland Ice Sheet. Preliminary attempts at applying the model with meltwater drainage to real-world scenarios have been performed in \citet{Woodsthesis}, and will be pursued in future work.

\backsection[Funding]%{Please provide details of the sources of financial support for all authors, including grant numbers. Where no specific funding has been provided for research, please provide the following statement: "This research received no specific grant from any funding agency, commercial or not-for-profit sectors." }
{Tilly Woods was supported by a doctoral studentship from the UK Engineering and Physical Sciences Research Council.}

\backsection[Declaration of interests]%{A Competing Interests statement is now mandatory in the manuscript PDF. Please note that if there are no conflicts of interest, the declaration in your PDF should read as follows: {\bf Declaration of Interests}. The authors report no conflict of interest.}
{The authors report no conflict of interest.}

\backsection[Data availability statement]{Upon publication of the manuscript, the code used to produce the final versions of the results and figures in this manuscript will be published in a Zenodo repository. For now, the current code is openly available at \hyperref[https://github.com/TillyWoods/unsat_WC_v2/tree/main]{https://github.com/TillyWoods/unsat\_WC\_v2/tree/main}. }
% See JFM's \href{https://www.cambridge.org/core/journals/journal-of-fluid-mechanics/information/journal-policies/research-transparency}{research transparency policy} for more information}

%\backsection[Author ORCIDs]{Authors may include the ORCID identifers as follows.  F. Smith, https://orcid.org/0000-0001-2345-6789; B. Jones, https://orcid.org/0000-0009-8765-4321}

%\backsection[Author contributions]{Authors may include details of the contributions made by each author to the manuscript'}

\backsection[Open Access]{For the purpose of open access, the author has applied a CC BY public copyright licence to any author accepted manuscript arising from this submission.}

\appendix

\section{Numerical method}
\label{sec4:numerical_method}

Here, we briefly outline the numerical method used to solve the model.  Further details can be found in \citet{Woodsthesis}.  

There are generally two out of the three variables $T$, $\phi$ and $S$ that must be solved for at any point in space and time, using equations \eqref{eq4:mass_ice_unsat_nondim}-\eqref{eq4:temp_eqn_unsat_nondim}.  In regions below the melting temperature $T_m$, there is no water so $S$ is undefined. In temperate regions the temperature $T = T_m$ is known, and both $\phi$ and $S$ must be solved for.  Energy and mass must be conserved at the (unknown) boundaries between these regions.  

A convenient method for solving this problem is the enthalpy method, which allows us to solve for a single set of variables (the enthalpy $H$ and total water content $W$) everywhere in the domain.  This removes the need to explicitly track the boundaries between the regions \citep{Woods2024}.

% The total water content $W$, defined as the sum of both the ice and liquid water contributions, is
% \be
% \label{eq4:W}
% W = 1 - \phi + S \phi,
% \ee
% and the enthalpy $H$, defined as the sum of the sensible and latent heat, is
% \be
% \label{eq4:H}
% H = \rho c W (T-T_{\mathrm{m}}) + \rho \mathcal{L} \phi S.
% \ee
% We can recover the porosity $\phi$, saturation $S$ and temperature~$T$ from the total water content $W$~\eqref{eq4:W} and the enthalpy $H$~\eqref{eq4:H} as:
% \be
% \begin{split}
% & \phi = \begin{cases}
% 1-W, & H \leq 0, \\
% 1-W + \frac{H}{\rho \mathcal{L}}, & 0<H<\rho \mathcal{L} W, \\
% 1, & H \geq \rho \mathcal{L} W ,\\
% \end{cases} \\
% & S = \begin{cases}
% 0, & H \leq0 ,\\
% \frac{H}{\rho \mathcal{L} \phi}, & 0<H< \rho \mathcal{L} W, \\
% W, & H \geq \rho \mathcal{L} W, \\
% \end{cases} \\
% & T = \begin{cases}
% T_m + \frac{H}{\rho c W}, & H \leq 0 ,\\
% T_m, & 0<H< \rho \mathcal{L} W ,\\
% T_m + \frac{H}{\rho c W} - \frac{\mathcal{L}}{c}, & H \geq \rho \mathcal{L} W.\\
% \end{cases}
% \end{split}
% \ee
% Here, the three cases $H<0$, $0<H< \rho \mathcal{L} W$ and $H> \rho \mathcal{L} W$ correspond to there being no water, both ice and water, and no ice, respectively. 

The total water content $W$ is defined as
\be
\label{eq4:W_nondim}
W = 1- \phi + \phi S,
\ee
and the dimensionless enthalpy $H$ is defined as
\be
\label{eq4:H_nondim}
H = WT + \mathcal{S} \phi S.
\ee
The porosity, saturation and temperature can be recovered from the enthalpy and total water content, as
\be
\label{eq4:phi_from_H}
\phi = \begin{cases}
1-W, & H \leq 0, \\
1-W + \frac{H}{\mathcal{S}}, & 0<H<\mathcal{S} W ,\\
1, & H \geq \mathcal{S} W, \\
\end{cases}
\ee
\be
\label{eq4:S_from_H}
S = \begin{cases}
0, & H \leq 0, \\
\frac{H}{\mathcal{S} \phi}, & 0<H< \mathcal{S} W, \\
W, & H \geq \mathcal{S} W, \\
\end{cases}
\ee
\be
\label{eq4:T_from_H}
T = \begin{cases}
\frac{H}{ W}, & H \leq0, \\
0, & 0<H< \mathcal{S} W, \\
\frac{H}{W} - \mathcal{S}, & H \geq \mathcal{S} W. \\
\end{cases}
\ee
Here, the three cases $H<0$, $0<H< \mathcal{S} W$ and $H> \mathcal{S} W$ correspond to there being no water, both ice and water, and no ice, respectively. 

%We are able to manipulate the model equations to express the problem in terms of $W$ and $H$ instead of $T$, $\phi$ and $S$. Summing the two mass conservation equations~\eqref{eq4:mass_ice_unsat_1d} and~\eqref{eq4:mass_water_unsat_1d}, we eliminate the melt rate $m_{\mathrm{int}}$ to get an equation for the total water content $W$,
We obtain equations for $W$ and $H$ by combining the mass conservation equations \eqref{eq4:mass_ice_unsat_nondim} and \eqref{eq4:mass_water_unsat_nondim} and the energy conservation equation \eqref{eq4:temp_eqn_unsat_nondim} to eliminate the subsurface melt rate $m\sub{int}$, giving
\be
\label{eq4:Weqn_nondim}
\pd{W}{t} + \pd{}{Z} \left( \dot{h} W + q \right) =  - q_{\mathrm{d}},
\ee
\be
\label{eq4:Heqn_nondim}
\pd{H}{t} + \pd{}{Z} \left( \dot{h} H -\hat{k} W \pd{T}{Z} + F + (\mathcal{S} + T) q \right) =  - (\mathcal{S}+T) q_{\mathrm{d}},
\ee

% \be
% \label{eq4:Weqn}
% \pd{W}{t} + \pd{}{Z} \left( \dot{h} W + q \right) = - \frac{q_{\mathrm{d}}}{L_{\mathrm{d}}}
% \ee
% and
% \be
% \label{eq4:Heqn}
% \begin{split}
% \pd{H}{t} + \pd{}{Z} \bigg( \dot{h} H - k W \pd{T}{z} + F + \left(\rho \mathcal{L} + \rho c (T-T_{\mathrm{m}}) \right) q \bigg) = &- \rho \mathcal{L} \frac{q_{\mathrm{d}}}{L_{\mathrm{d}}}\\
% &- \rho c (T - T_{\mathrm{m}} ) \frac{q_{\mathrm{d}}}{L_{\mathrm{d}}}.
% \end{split}
% \ee

% Furthermore, combining the two kinematic conditions~\eqref{eq4:kinematic_water_nondim} and~\eqref{eq4:kinematic_ice_nondim} to eliminate the surface melt rate $m_{\mathrm{surf}}$ gives a surface boundary condition in terms of $W$,
% \be
% \label{eq4:kinematic_combined_nondim}
% q +\dot{h} W = R_{\mathrm{r}} - r_{\mathrm{off}} \quad \mathrm{at} \quad Z=0.
% \ee

We solve the enthalpy equation~\eqref{eq4:Heqn_nondim} and total water content equation~\eqref{eq4:Weqn_nondim} using a finite volume scheme with semi-implicit timestepping in MATLAB. A delicate aspect of the method is that the water pressure $p_{\mathrm{w}}$ and flux $q$ must be calculated differently in saturated ($S=1$) and unsaturated ($0<S<1$) regions.  

In unsaturated regions, $p_{\mathrm{w}}$ is parametrised in terms of saturation (given by~\eqref{eq4:pw_nondim}) and so the expression~\eqref{eq4:Darcy_nondim} for $q$ becomes
\be
\label{eq4:unsat_flux_nondim}
q = \mathcal{K} \phi^3 \left( S^3 - \frac{2}{B} \pd{S}{Z} \right) \quad \mathrm{when} \quad 0<S<1.
\ee

In saturated regions, the flux becomes
\be
\label{eq4:sat_flux_nondim}
q = \mathcal{K} \phi^3 \left( 1 - \pd{p_{\mathrm{w}}}{Z} \right) \quad \mathrm{when} \quad S=1,
\ee
and satisfies (from~\eqref{eq4:Weqn_nondim} with $W=1$)
\be
\label{eq4:q_eqn_sat}
\begin{split}
    \pd{q}{Z} &= -q_{\mathrm{d}},
\end{split}
\ee
where $q_{\mathrm{d}}= \mathcal{K}_{\mathrm{d}} \mathcal{K} \phi^3$ when $S=1$. Hence the water pressure $p_{\mathrm{w}}$ satisfies the elliptic equation
\be
\label{eq4:pw_eqn_nondim_early}
\pd{}{Z} \bigg( \mathcal{K} \phi^3 \Big( 1 - \pd{p_{\mathrm{w}}}{Z} \Big) \bigg) = -\mathcal{K}_{\mathrm{d}} \mathcal{K} \phi^3.
\ee

In saturated regions without a solid base, we solve equation~\eqref{eq4:pw_eqn_nondim_early} for $p_{\mathrm{w}}$. The boundary conditions are
continuity with atmospheric pressure $p_{\mathrm{w}}=0$ at the surface and at interfaces between saturated and unsaturated regions, and no flux at impermeable, solid boundaries (where $\phi = 0$). The flux $q$ is then calculated using equation~\eqref{eq4:sat_flux_nondim}. 
(We apply $p_{\mathrm{w}}=0$ at saturated--unsaturated interfaces, instead of continuity with the unsaturated pressure $p_{\mathrm{w}} = -p_c(S) = - 1/({\mathcal{B}}S^{2})$, to avoid numerical issues when there are large unresolved gradients in $S$ at these interfaces, and note that the two conditions are equivalent when $\mathcal{B} \to \infty$.)

In saturated regions with a solid base (e.g. the bottom of the weathering crust), the no-flux condition at the impermeable base allows us to integrate the $q$ equation~\eqref{eq4:q_eqn_sat} directly to find $q$, without ever needing to solve for $p_{\mathrm{w}}$. In our numerical method, we do this for saturated regions with a solid-ice base. In fact, in saturated regions bounded on both sides by solid ice, the $p_{\mathrm{w}}$ equation is ill-posed (the pressure is only determined there up to a constant).

\section{Semi-analytical solution for isothermal water transport}
\label{app:isothermal_water}

In this appendix, we present the method used to find the semi-analytical solution to the isothermal water transport test case presented in Section~\ref{sec4:isothermal_water}.

\subsection{Wetting fronts --- before saturation occurs}
In the initial phase before saturation occurs, we only need to solve the unsaturated form of equation~\eqref{eq4:unsaturated}, which it is helpful to express in terms of a new variable $\tilde{W} = \phi S$, as
\be
\label{eq4:Wtilde_eqn_noncon}
\pd{\tilde{W}}{t} + 3 \mathcal{K} \tilde{W}^2 \pd{\tilde{W}}{Z} = - \mathcal{K}_{\mathrm{d}} \mathcal{K} \tilde{W}^{3}.
\ee
This can be solved using the method of characteristics. The characteristic equations are
\be
\dot{t}=1, \quad \dot{Z} = 3 \mathcal{K} \tilde{W}^2, \quad \dot{\tilde{W}} = - \mathcal{K}_{\mathrm{d}} \mathcal{K} \tilde{W}^{3},
\ee
where dot denotes the derivative along the characteristics.
Along the characteristics,
\be
\dd{\tilde{W}}{Z} = \frac{\dot{\tilde{W}}}{\dot{Z}} = -\frac{\mathcal{K}_{\mathrm{d}}}{3} \tilde{W},
\ee
and so
\be
\tilde{W} = A e^{-\mathcal{K}_{\mathrm{d}} Z/3},
\ee
where $A$ is constant along each characteristic.

There are two sets of characteristics: one set from $t=0$ (to which we apply the initial condition~\eqref{eq4:isothermal_initial}) and one set from $Z=0$ (to which we apply the boundary condition~\eqref{eq4:isothermal_boundary}). Our initial condition~\eqref{eq4:isothermal_initial} is expressed in terms of $\tilde{W}$ as
\be
\tilde{W} = 0 \quad \mathrm{at} \quad t=0.
\ee
Hence, on the characteristics starting from $t=0$, we have $A=0$. The boundary condition~\eqref{eq4:isothermal_boundary} for constant flux (rainfall) at the surface can be expressed in terms of $\tilde{W}$ as
\be
\tilde{W} = \bigg( \frac{\mathcal{R}_{\mathrm{r}}}{\mathcal{K}} \bigg)^{1/3} \quad \mathrm{at} \quad Z=0.
\ee
Hence, the characteristics starting from $Z=0$ have $A = (\mathcal{R}_{\mathrm{r}} / \mathcal{K} )^{1/3}$.

The two sets of characteristics intersect (creating a `shock') at a wetting front ${Z=Z_{\mathrm{w}}(t)}$, the depth to which water from the surface has infiltrated, with the region below this being dry. The position of $Z_{\mathrm{w}}(t)$ is found using the Rankine-Hugoniot condition \citep{Ockendon2016}, giving
\be
\label{eq4:Zw_RH_cond}
\dot{Z}_{\mathrm{w}} = \frac{[\mathcal{K} \tilde{W}^3]_{-}^{+}}{[\tilde{W}]_{-}^{+}} = \mathcal{K} \tilde{W}_{-}^{2},
\ee
where the square brackets represent the jump in the solution across $Z_{\mathrm{w}}$, with `$-$' being the solution from the $Z=0$ characteristics, and `$+$' being the solution from the $t=0$ characteristics.

Combining the above expressions, we find the solution
\be
\label{eq4:Wtilde_ana_time_evol}
\tilde{W} =
\begin{cases}
    \big( \frac{\mathcal{R}_{\mathrm{r}}}{\mathcal{K}} \big)^{1/3} e^{-\mathcal{K}_{\mathrm{d}} Z/3}, & Z<Z_{\mathrm{w}}(t), \\
    0, & Z>Z_{\mathrm{w}}(t),
\end{cases}
\ee
from which we recover solutions \eqref{eq4:S_ana_time_evol} and \eqref{eq4:q_ana_time_evol} for $S$ and $q$, with the wetting front $Z\sub{w}(t)$ given by equation \eqref{eq4:Zw_drainage}.
This solution holds until saturation ($S=1$) first occurs.

% For sufficiently small lateral drainage ($\mathcal{K}_{\mathrm{d}}<3/Z_0$), the saturation increases with depth, with $S=1$ first being reached at the wetting front after a time $t_{\mathrm{s}}$. The position $Z_{\mathrm{s}}$ at which this occurs is found by setting $S(Z_{\mathrm{s}})=1$, giving
% \be
% \label{eq4:Zs_drainage}
% Z_{\mathrm{s}} = \frac{Z_0}{3 - Z_0 \mathcal{K}_{\mathrm{d}}} \log \Bigg( \frac{\mathcal{K} \phi_0^3 }{q_0} \Bigg).
% \ee
% The time $t_{\mathrm{s}}$ at which saturation first occurs is then found by equating $Z_{\mathrm{s}}$ to the wetting from position $Z_{\mathrm{w}}$~\eqref{eq4:Zw_drainage}, resulting in
% \be
% \label{eq4:ts_drainage}
% t_{\mathrm{s}} = \frac{3}{2 \mathcal{K}_{\mathrm{d}} \mathcal{K}^{1/3} q_0^{2/3}} \Bigg( \bigg( \frac{\mathcal{K} \phi_0^3}{q_0} \bigg)^{\frac{2 Z_0 \mathcal{K}_{\mathrm{d}}}{9 - 3 Z_0 \mathcal{K}_{\mathrm{d}}}} - 1 \Bigg).
% \ee
% For this solution to hold, we must have $\mathcal{K}_{\mathrm{d}}<3/Z_0$. That is, the lateral drainage must not be too large. If $\mathcal{K}_{\mathrm{d}}>3/Z_0$, the saturation decreases with depth and never reaches 1.

\subsection{Saturation fronts --- after saturation first occurs}
Once saturation first occurs at $Z=Z_{\mathrm{s}}$ at time $t=t_{\mathrm{s}}$, we have a growing saturated region $Z_{\mathrm{t}} (t) < Z < Z_{\mathrm{b}} (t)$ in which $S=1$. The top boundary $Z_{\mathrm{t}}(t)$ moves upwards (decreases in value, given the orientation of $Z$) and the bottom boundary $Z_{\mathrm{b}}(t)$ moves downwards (increases in value). We can find these saturation fronts by solving the equations~\eqref{eq4:saturated} for the water flux $q$ and and water pressure $p_{\mathrm{w}}$ in the saturated region. This tells us that the flux $q$ in the saturated region is
\be
q = q_{\mathrm{s}} + \frac{Z_0}{3} \mathcal{K}_{\mathrm{d}} \mathcal{K} \phi^3 \quad \mathrm{for} \quad Z_{\mathrm{t}}(t)<Z<Z_{\mathrm{b}}(t), \ \ t>t_{\mathrm{s}},
\ee
for some $q_{\mathrm{s}}(t)$ independent of $Z$. The value of $q_{\mathrm{s}}$ is found by integrating the expression in equation~\eqref{eq4:saturated} for the flux in the saturated region from $Z = Z_{\mathrm{t}}$ to $Z=Z_{\mathrm{b}}$. This tells us that
\be
p_{\mathrm{w}} (Z_{\mathrm{b}}) - p_{\mathrm{w}} (Z_{\mathrm{t}}) = Z_{\mathrm{b}} - Z_{\mathrm{t}} - \frac{q_{\mathrm{s}}}{\mathcal{K}} \int_{Z_{\mathrm{t}}}^{Z_{\mathrm{b}}} \frac{ e^{-\mathcal{K}_{\mathrm{d}}Z}}{\phi(Z)^3} \mathrm{d}Z.
\ee
We require the water pressure $p_{\mathrm{w}}$ to be continuous between saturated and unsaturated regions. In the unsaturated region, $p_{\mathrm{w}} =0$ (since we are neglecting capillary pressure), so we have that $p_{\mathrm{w}} (Z_{\mathrm{b}}) = p_{\mathrm{w}} (Z_{\mathrm{t}}) =0$. Evaluating the integral tells us that $q\sub{s}$ is given by equation \eqref{eq4:qs0}.

To determine equations for the saturation fronts, we use jump conditions that must be satisfied at the fronts in order to conserve mass. These are found by integrating the mass conservation equation~\eqref{eq4:unsaturated} across each front. This gives
\be
0 = [q-\phi S \dot{Z}_{\mathrm{t}}]_{-}^{+}, \quad 0 = [q-\phi S \dot{Z}_{\mathrm{b}}]_{-}^{+},
\ee
where, as previously, `$+$' indicates above the front and `$-$' indicates below. At $Z=Z_{\mathrm{b}}$, we have $q_- = 0$, $S_- = 0$ (no water below the front) and $q_+ = q_{\mathrm{s}} $, $S_+ = 1$ (saturated above the front). At $Z=Z_{\mathrm{t}}$, we have $q_- = q_{\mathrm{s}} $, $S_- = 1$ (saturated below) and $q_+$, $S_+$ are given by the unsaturated solutions~\eqref{eq4:q_ana_time_evol},~\eqref{eq4:S_ana_time_evol}. Substituting all of this in gives the two coupled ordinary differential equations \eqref{eq4:sat_front_ODEs_drainage}
% \be
% \label{eq4:sat_front_ODEs_drainage}
% \dot{Z}_{\mathrm{t}} = -\frac{q_0 e^{-\mathcal{K}_{\mathrm{d}} Z_{\mathrm{t}}}-\big(q_{\mathrm{s0}} + \frac{Z_0}{3} \mathcal{K}_{\mathrm{d}} \mathcal{K} \phi_{\mathrm{t}}^3 \big)}{\phi_{\mathrm{t}} - (q_0/\mathcal{K})^{1/3} e^{-\mathcal{K}_{\mathrm{d}} Z_{\mathrm{t}}/3}}, \quad \dot{Z}_{\mathrm{b}} = \frac{q_{\mathrm{s0}} + \frac{Z_0}{3} \mathcal{K}_{\mathrm{d}} \mathcal{K} \phi_{\mathrm{b}}^3}{\phi_{\mathrm{b}}},
% \ee
for $Z_{\mathrm{t}}(t)$ and $Z_{\mathrm{b}}(t)$.
% , where $\phi_{\mathrm{t}} = \phi (Z_{\mathrm{t}})$, $\phi_{\mathrm{b}} = \phi (Z_{\mathrm{b}})$.
We solve the differential equations~\eqref{eq4:sat_front_ODEs_drainage} for the saturation fronts numerically in MATLAB using \texttt{ode15s}, with initial conditions  $Z_{\mathrm{t}} = Z_{\mathrm{b}} = Z_s$ at $t=t_s$.

\bibliography{thesis_references}
\bibliographystyle{jfm}

\end{document}